\documentclass[onecolumn, trackchanges]{aastex7}

\usepackage{amsmath}
\usepackage{booktabs}
\usepackage{comment}
\usepackage[utf8]{inputenc}
\begin{document}

\title{Beyond Extreme Burstiness: Evolving Star Formation Efficiency as the Key to Early Galaxy Abundance}

\correspondingauthor{Abhijnan Kar}

\author{Abhijnan Kar}
\affiliation{Department of Physical Sciences, Indian Institute of Science Education and Research Berhampur, Vigyanpuri, Ganjam, 760003, India}
\email[show]{karabhijnan123@gmail.com}  

\author{Shadab Alam}
\affiliation{Tata Institute of Fundamental Research, Homi Bhabha Road, Mumbai 400005, India}
\email[show]{shadab@theory.tifr.res.in}

\author{Joseph Silk}
\affiliation{William H. Miller III Department of Physics and Astronomy, The Johns Hopkins University, Baltimore, Maryland 21218, USA}
\affiliation{Institut d’Astrophysique de Paris, UMR 7095 CNRS and UPMC, Sorbonne Universite´, F-75014 Paris, France}
\email[show]{silk@iap.fr}

\begin{abstract}

JWST observations have revealed an overabundance of bright galaxies at $z \geq 9$, creating apparent tensions with theoretical predictions within standard $\Lambda$CDM cosmology. We address this challenge using a semi-empirical approach that connects dark matter halos to the  observed UV luminosity through physically motivated double power-law star formation efficiency (SFE) model as a function of halo mass, redshift and perform joint Bayesian analysis of luminosity functions spanning $z = 4-16$ using combined HST and JWST data. Through systematic model comparison using information criteria (AIC, BIC, DIC), we identify the optimal framework requiring redshift evolution only in the low-mass slope parameter $\alpha(z)$ while maintaining other SFE parameters constant. Our best-fitting model achieves excellent agreement with observations using modest, constant UV scatter $\sigma_{\rm UV} = 0.32$ dex—significantly lower than the $\gtrsim 1.3$ dex values suggested by previous studies for $z > 13$. This reduced scatter requirement is compensated by strongly evolving star formation efficiency, with $\alpha$ increasing toward higher redshifts, indicating enhanced star formation in low-mass halos during cosmic dawn. The model also successfully reproduces other important observational diagnostics such as  effective galaxy bias and cosmic Star Formation Density (SFRD) consistently across the full redshift range. Furthermore, model predictions are consistent up to a redshift of $z\sim 20$. Our results demonstrate that JWST's early galaxy observations can be reconciled with standard cosmology through the interplay of modest stochasticity and evolving star formation physics, without invoking extreme burstiness or exotic mechanisms. 
\end{abstract}



\section{Introduction} 

The James Webb Space Telescope (JWST) has ushered in a new era in understanding the formation and evolution of galaxies during the initial ~500 million years of cosmic history, especially at redshifts \( z \gtrsim 10 \). In continuation with earlier HST data \citep[upto $z\sim9$]{Oesch2018,Bouwens2021, Bouwens2014}, an increasing number of bright galaxies have been found via photometric detection (e.g., \citet{Naidu2022b,Castellano2022,Finkelstein2022,Adams2023,Atek2023,Bouwens2023b,Harikane2023}. Furthermore, this includes very bright sources located at redshifts upto $z\sim 19$ (e.g., \citet{Harikane2023,perez-gonzalez_notitle_2025,whitler2025}) some of which have been spectroscopically validated through various JWST surveys \citep{
%naidu_notitle_2025,
Curtis2023,Carniani2024,Fujimoto2023b,Castellano2024}.
The highest spectroscopically confirmed redshift galaxy is at $z=14.44$
\citep{naidu_notitle_2025}.
The UV Luminosity Functions (LFs) from these observations  show little evolution at the brighter end \citep{Harikane2023} and the number densities of these bright galaxies often exceed the theoretically predicted LFs from pre JWST models \citep[e.g,]{Finkelstein2022a}. Several different approaches have been used to model such number densities earlier from empirical \citep{Tacchella2013,Mason2015,Sun2016,Behroozi2020}, semi-analytical models \citep[][ and so on]{Dayal2014, Dayal2019} to hydrodynamical simulations \citep{Vogelsberger2020, Feldmann2024, Flores2021, Wilkins2023, Wilkins2023b}. Of late, \citet{donnan_no_2025} also proposed the idea of young stellar ages at high redshift as a factor contributing to the UVLF instead of an increased star formation efficiency.

Different explanations ranging from cosmological modifications to astrophysical aspects have been proposed in order to reconcile such observations with theoretical predictions. Some works such as \citet{koehler_notitle_2024, ShenEDE} explore modifications of $\Lambda\rm C\rm D\rm M$ as a solution to achieve the enhanced galaxy number densities. However, whether such a radical option is merited is debatable, and several attempts have been made towards understanding these observations within the standard cosmology framework.  One such possibility is increased UV flux from these galaxies through astrophysical processes such as enhanced Star Formation Efficiency (SFE) \citep{Dekel2023,Mason2023, Li2023, nikopoulos_notitle_2024} or via a top-heavy stellar Initial Mass Function (IMF) \citep{Inayoshi2022,Yung2023,Cueto2023, Trinca2024, Ventura2024, mauerhofer_notitle_2025}, or even by invoking  early dust attenuation \citep{Ferrara2023,toyouchi_notitle_2025}.

Along with the explanations using enhanced UV luminosity, another possibility could be the variability in observed UV luminosity, represented by a distribution with scatter $\sigma_{\rm UV}$ \citep{Shen2023,sun_notitle_2023,Mirocha2023, Kravtsov2024}. Such variability or stochasticity could introduce more up-scatter of LF at the brighter end compared to the dimmer end, through Eddington Bias \citep{eddington}. One of the most plausible reasons behind such variability could be stochastic or bursty star formation scenario in such galaxies at high redshift. Different mechanisms such as mergers and feedback processes are thought to contribute to this bursty star formation history (SFH). This has also been found in several zoom-in hydrodynamical simulations \citep{Hopkins2018,Marinacci2019,sun_notitle_2023,Katz2023,Flore2023}. Moreover, recent observational studies have shed light into the burstiness in the star formation histories of high-redshift galaxies \citep{Ciesla2023,Cole2023,Endsley2023,Looser2023,Tacchella2023, Dressler2024, Helton2024, kokorev_capers_2025}, also suggesting a possible significant change in star formation patterns around $z \sim 10$. Such bursty star formation histories create interesting scenarios for further observations too e.g, "quiescent" galaxies in post-starburst mode \citep{gelli_notitle_2025}. Thus, understanding these aspects of star formation in high redshift galaxies is necessary in order to understand the evolution of galaxy number densities.\\
Different simulations and semi-empirical works have tried to quantify the scatter $\sigma_{\rm UV}$. Works from FIRE-2, FIREBox and SPHINX simulations \citep{sun_notitle_2023,Feldmann2024,Katz2023} show a maximum possible scatter around 2.0 dex at these redshifts, where as, from SERRA simulations, the value of scatter goes as low as $\sim 0.6$ dex \citep{Ferrara2023}. On the contrary, different semi-empirical and analytical works highlight the need for very high scatter values possibly breaching any sort of maximum limit. For example, analysis in \citet{Shen2023} requires a very high consistent scatter of ($\sigma_{\rm UV} \sim 1.5–2.5$) dex to match LFs around ( $z \sim 12–16 $), while more detailed modeling of bursty star-formation histories suggests a slightly reduced ($\sigma_{\rm UV} \sim 1–1.3$) mag at ( $z \sim 12 $) \citep{Kravtsov2024}. Further, \citet{Gelli2024} introduces a halo mass-dependent scatter model, which is effective in describing observation only up to $z\sim 13$. Moreover, most of these semi-empirical works used ad hoc values in crucial astrophysical parameters e.g, SFE etc \citep[see][]{ShenEDE} which were based on earlier pre-JWST calibrations or models.
Hence, it becomes important to understand contributions from both enhanced SFE and burstiness causing UV variability in a data-driven manner. Some previous works such as \citet{sipple_notitle_2024} looked into the effect of scatter and star formation while fitting the joint LF within a Bayesian sampling framework. However, their joint LF fitting used only HST data upto $z \sim 8$. Indeed, this way of getting the constraint on such parameters and their redshift evolution (if any) helps in drawing a self-consistent picture from the data. More recently, \citet{shuntov2025sigma} reports a constant modest scatter value ($\sim 0.6$) needed to account for the LFs along with SFE evolution, from the limited FRESCO survey data upto $z \sim 8$. This hints towards potential degeneracy between SFE and scatter in UV luminosity as well as their redshift evolution which should be explored within a  self-consistent framework.
This trade-off or balance between SFE evolution and $\sigma_{\rm UV}$ could reproduce the high number densities without the need for an exotic cosmological scenario. Recent works by \citet{somerville_density_2025, dhandha_exploiting_2025} also explore such evolving SFEs and requirement of bursty star formation in explaining UVLFs at high redshift, albeit within a different modelling framework.

In this paper,  we consider a consistent Bayesian framework simultaneously analyzing UV luminosity function of galaxies from redshift 4 to 16 using HST and JWST observations. We also perform Bayesian model selection to understand the models with minimal freedom which can explain the observations over this wide range of cosmic evolution. We also study the necessity of any redshift evolution of different parameters (e.g., SFE slopes, scatter), to explain the observations within $\Lambda$CDM. We describe the semi-empirical model (based on \citet{ShenEDE, Shen2023}) used in section \ref{methods}. In Section \ref{Results}, the best-describing model and its redshift dependent parametrization are discussed along with the evolution of different astrophysical parameters from the model. We discuss implications of our results and provide a key summary in Section \ref{sec: discussion}. Throughout this work, we have assumed flat-$\Lambda$CDM cosmology with parameter values from Planck measurements \citep{Planck2020}, unless stated otherwise.

\section{Methods}\label{methods}
\subsection{Semi-Empirical Model of Luminosity Function}\label{sec: theory StandardLF}

To construct a physically-motivated galaxy luminosity function at high redshifts, our approach connects dark matter halos to their host galaxies through empirical star formation prescriptions. This semi-empirical framework allows us to predict observable galaxy properties from the underlying dark matter structure.

\subsubsection{Dark Matter Foundation: Halo Mass Function}
We begin with the dark matter halo mass function (HMF), which describes the comoving number density of halos per logarithmic mass interval:
\begin{equation}
    \frac{dn}{d\log M_{\rm H}} = f(\sigma) \frac{\overline{\rho}_{\rm m}}{M_{\rm H}} \frac{d\ln\sigma^{-1}}{d\ln M_{\rm H}}
\end{equation}
where $f(\sigma)$ is the halo multiplicity function and $\sigma$ is the rms fluctuation on the scale of halo mass $M_{\rm H}$. We use the calibration from \citet{Tinker2008}, implemented in the HMF package \citep{hmf3,hmf2}. Other HMF calibrations from \citet{Behroozi2013, rodriguez-puebla_halo_2016}, Shin-Uchuu simulation suites \citep{Ishiyama2021} have only small offset wrt \citet{Tinker2008}, by a factor between 1.2 - 1.5 (Fig \ref{HMF_compare}). Our halo mass range spans $10^8$ to $10^{13}\,\mathrm{M}_{\odot}$, encompassing the hosts of faint to bright early galaxies at redshifts $z = [4,5,6,7,8,9,11,12.5,14,16]$.

\subsubsection{Connecting Halos to Galaxies: The $M_{\rm UV} - M_{\rm H}$ Relation}
The key physical ingredient is translating halo mass into observed UV luminosity through star formation. We construct this relation via:
\begin{equation}
    M_{\rm UV} \leftarrow L_{\rm UV} \leftarrow \mathrm{SFR} \leftarrow \epsilon(M_{\rm H}) \cdot \dot{M}_{\rm H}
\end{equation}

The star formation efficiency (SFE) $\epsilon(M_{\rm H})$ determines what fraction of the accreting gas forms stars. Following previous works \citep{Shen2023, ShenEDE, Tacchella2018, Moster2010, Harikane2022}, we adopt a double power-law form:
\begin{equation}
    \epsilon(M_{\rm H}) = \frac{2\epsilon_{0}}{(M_{\rm H}/M_{0})^{-\alpha} + (M_{\rm H}/M_{0})^{\beta}}
    \label{SFE}
\end{equation}
This functional form captures the physical expectation that star formation is suppressed in both low-mass halos (due to photoheating and supernova feedback) and high-mass halos (due to virial shock heating and AGN feedback). The parameters are:
\begin{itemize}
    \item $\epsilon_{0}$: Maximum SFE amplitude ($2\epsilon_{0} \leq 1$ maintains $\mathrm{SFR} \leq \dot{M}_{\rm H}$)
    \item $M_{0}$: Characteristic halo mass where SFE transitions from low to high mass behavior
    \item $\alpha$: Low-mass slope (steeper values indicate stronger feedback)
    \item $\beta$: High-mass slope (controls AGN/virial shock suppression)
\end{itemize}

\subsubsection{Mass Accretion Rate}
The halo mass accretion rate $\dot{M}_{\rm H}$ is obtained from the fitting formula of \citet{RP2016}, calibrated against the Bolshoi-Planck and Multidark-Planck simulations \citep{Klypin2016}:
\begin{equation}
    \dot{M}_{\rm H} = C \left(\frac{M_{\rm H}}{10^{12}\,\mathrm{M}_{\odot}/h}\right)^{\gamma} \frac{H(z)}{H_0}
    \label{halo accretion}
\end{equation}
where the redshift-dependent parameters are:
\begin{align}
    \gamma &= 1.000 + 0.329\,a - 0.206\,a^2 \\
    \log_{10} C &= 2.730 - 1.828\,a + 0.654\,a^2
\end{align}
and $a = 1/(1+z)$ is the scale factor. This fitting formula captures the physical trend that higher-mass halos accrete more rapidly, with the rate increasing toward higher redshifts when halos grow more vigorously.

\subsubsection{From Star Formation to UV Luminosity}
The star formation rate is computed as:
\begin{equation}
    \mathrm{SFR} = \epsilon(M_{\rm H}) \cdot f_{\rm b} \cdot \dot{M}_{\rm H}
\end{equation}
where $f_{\rm b} = \Omega_{\rm b}/\Omega_{\rm m}$ is the universal baryon fraction. This assumes that $\epsilon \cdot \dot{M}_{\rm H}$ of the accreting baryonic matter converts to stars.

We convert SFR to UV luminosity using the relationship calibrated for a \citet{Chabrier2003} initial mass function:
\begin{equation}
    \mathrm{SFR} = \kappa \cdot L_{\nu}(\lambda = 1500\,\text{Å})
\end{equation}
where $\kappa = 1.15 \times 10^{-28}$ (in units of $\mathrm{M}_{\odot}\,\mathrm{yr}^{-1}$ per $\mathrm{erg}\,\mathrm{s}^{-1}\,\mathrm{Hz}^{-1}$). However, the value of $\kappa$ is dependent on metallicity considered and stellar age which in turn depends on the redshift assumed as at high redshift there would be less time to form all stars \citep{Harikane2023, donnan_no_2025}. The absolute UV magnitude is then:
\begin{equation}
    M_{\rm UV} = -2.5\log_{10}\left(\frac{L_{\nu}}{4\pi d^2}\right) - 48.6
\end{equation}
where $d = 10\,\mathrm{pc}$ by definition of absolute magnitude.

\subsubsection{Constructing the Luminosity Function}
Now, once we have a $M_{\rm UV} - M_{\rm H}$ relation, the underlying luminosity function is obtained by transforming the halo mass function:
\begin{equation}
    \frac{dn}{dM_{\rm UV}} = \frac{dn}{d\log M_{\rm H}} \cdot \left \lvert \frac{d\log M_{\rm H}}{dM_{\rm UV}} \right \rvert
    \label{LF eqn}
\end{equation}

\subsubsection{Stochasticity and Scatter}
Real galaxies exhibit scatter in their star formation histories, leading to variations in UV luminosity at fixed halo mass. We model this stochasticity by introducing a log-normal halo mass independent scatter $\sigma_{\rm UV}$ in the $M_{\rm UV} - M_{\rm H}$ relation. This scatter creates an Eddington bias \citep{eddington}, preferentially scattering galaxies upward in luminosity at the bright end of the function.

We account for this by convolving the underlying luminosity function with a Gaussian kernel:
\begin{equation}
    \Phi_{\rm obs}(M_{\rm UV}) = \int_{-\infty}^{\infty} \Phi_{\rm o}(M'_{\rm UV}) \cdot \frac{1}{\sqrt{2\pi}\sigma_{\rm UV}} \exp\left[-\frac{(M_{\rm UV} - M'_{\rm UV})^2}{2\sigma_{\rm UV}^2}\right] dM'_{\rm UV}
\end{equation}

\subsubsection{Dust Attenuation}
We incorporate dust extinction using empirical relations. The UV attenuation is related to the UV slope $\beta$ through \citet{Meurer1999}:
\begin{equation}
    A_{\rm UV} = 4.43 + 1.99\beta
\end{equation}
The slope-magnitude relation follows \citet{Cullen2023} for $8 \leq z \leq 10$:
\begin{equation}
    \beta = -0.17M_{\rm UV} - 5.40
\end{equation}
For $z < 8$, we use the relation from \citet{Bouwens2014}. At $z > 10$, we assume negligible dust attenuation, consistent with the expectation of minimal dust in the earliest galaxies \citep{Cullen2024}.
\subsubsection{Star Formation Rate Density}
The Star Formation Rate Density (SFRD) for each redshift is calculated using UV luminosity density ($\rho_{\rm UV}$) as follows:
\begin{equation}
    \rho_{\rm UV} = \int_{L(M_{\rm UV,\rm limit})}^{\infty}dL~L\Phi(L,z)
\end{equation}
We take the $M_{\rm UV,\rm limit}$ to be $-13$ and $-17$ in two different cases, while the upper limit is taken to be luminosity at the brightest end $M_{\rm UV}$. Now SFRD is obtained:
\begin{equation}
    \rho_{\rm SFR} = \kappa ~\rho_{\rm UV}
\end{equation}
where $\kappa$ value is taken to be $1.15\times10^{-28}$ $\rm M_{\odot}\rm yr^{-1}~\rm per~ \rm erg~s^{-1}~Hz^{-1}$.
\subsubsection{Galaxy Bias}
We calculate the luminosity-weighted effective bias following \citet{Gelli2024, Munoz2023}:
\begin{equation}
    b_{\rm eff}(M_{\rm UV}) = \frac{1}{\Phi(M_{\rm UV})} \int dM_{\rm H} \, P(M_{\rm UV}|M_{\rm H}) \, \frac{dn}{dM_{\rm H}} \, b(M_{\rm H})
    \label{eq: bias}
\end{equation}
where the conditional probability incorporates scatter:
\begin{equation}
    P(M_{\rm UV}|M_{\rm H}) = \frac{1}{\sqrt{2\pi}\sigma_{\rm UV}} \exp\left[-\frac{[M_{\rm UV} - M_{\rm UV,c}(M_{\rm H},z)]^2}{2\sigma_{\rm UV}^2}\right]
    \label{eq: PDF}
\end{equation}
Here, $M_{\rm UV,c}(M_{\rm H},z)$ is the central $M_{\rm UV} - M_{\rm H}$ relation from our best-fit parameters, and $b(M_{\rm H})$ is the halo bias from \citet{Tinker2010} as implemented in the \texttt{COLOSSUS} package \citep{Diemer2018}. By integrating over the bias values weighted by $\Phi$ upto a $M_{\rm UV}$ value (upper limit) with appropiate normalization factor, we calculate integrated bias as a function of redshift. 

\subsection{Redshift Evolution and Bayesian Parameter Estimation}\label{MCMC}

A critical aspect of our model is understanding how galaxy formation physics evolves with cosmic time. Previous studies \citep{Shen2023, ShenEDE} often assumed fixed star formation efficiency parameters based on simulations or pre-JWST SED fitting. However, to provide the most robust constraints, we allow our key parameters to evolve with redshift and directly constrain them from the observed luminosity function data itself.

\subsubsection{Parametric Redshift Evolution}\label{redshift evol}
We model the redshift dependence of our key parameters using either polynomial or power-law forms. Each parameter $P$ can evolve as:
\begin{align}
    P(z) &= P_0 + P_1 z + P_2 z^2 \quad \text{(polynomial form)}\label{eq:polynomial} \\
    P(z) &= P_0 (1+z)^r + c \quad \text{(power-law form)}\label{eq:powerlaw}
\end{align}

We consider redshift evolution for:
\begin{itemize}
    \item $\alpha(z)$: Low-mass slope of star formation efficiency
    \item $\beta(z)$: High-mass slope of star formation efficiency
    \item $\epsilon_0(z)$: Maximum star formation efficiency amplitude
    \item $M_0(z)$: Characteristic halo mass (evolved in log-space: $\log M_0 = M_1 + M_2 z + M_3 z^2$)
    \item $\sigma_{\rm UV}(z)$: UV scatter parameter
\end{itemize}

A redshift-dependent SFE essentially changes the underlying luminosity function shape with cosmic time, potentially capturing the evolving physics of feedback, gas accretion, and stellar mass assembly. A higher SFE at high redshift could potentially compensate for a low scatter, thereby, requiring only less UV scatter than earlier expectation.
Alongside, we also consider a scenario where redshift dependence kicks in for $z>10$:
\begin{equation*}
\mathrm{SFE}(z) =
\begin{cases}
\text{$z$-independent}, & z < 10, \\[6pt]
z\text{-dependent } ;\alpha(z), & z \geq 10.
\end{cases}
\end{equation*}

While such SFE parametrization can have discrete or sharp transition at $z = 10$, re-parametrization of $\alpha(z)$ as $\alpha(z -10)$ $\forall$ $z>10$ provides a smooth transition around $z = 10$ and it is useful for comparison between models.

\subsubsection{Bayesian Analysis with MCMC}
We employ Markov Chain Monte Carlo (MCMC) to sample the posterior probability distribution of our parameters, allowing us to quantify uncertainties and correlations. Our likelihood function assumes Gaussian errors:
\begin{equation}
    \chi^2 = -2\ln\mathcal{L} = \sum_{i,z} \frac{[\phi_{\rm model}(M_{\rm UV,i}, z) - \phi_{\rm obs}(M_{\rm UV,i}, z)]^2}{\sigma_i^2}
\end{equation}
where the sum extends over all magnitude bins $i$ and redshift bins $z$, and $\sigma_i$ represents the combined measurement uncertainty.

We implement the MCMC sampling using the \texttt{emcee} package, with post-processing handled by the publicly available \texttt{EASYmcmc} framework\footnote{\url{https://gitlab.com/shadaba/easymcmc}}. Convergence is assessed using the Gelman-Rubin criterion $\hat{R} \sim 1.0$ \citep{gelman-rubin}. For e.g, our best fit model shows an average $\hat{R} = 1.002$ for the free parameters in the diagnostic criterion.

\subsubsection{Prior Selection}
We adopt uniform priors for all parameters, with ranges chosen to be physically reasonable while allowing sufficient freedom for the data to constrain the values. Key constraints include:
\begin{itemize}
    \item Maximum star formation efficiency: $2\epsilon_0 \leq 1$ (no galaxy can convert more than entire gas to stars)
    \item UV scatter: $\sigma_{\rm UV} \leq 2.0$ dex (following \citealt{ShenEDE, Feldmann2024, Kravtsov2024})
    \item Slope parameters: Constrained to avoid sign flips that would be unphysical.
\end{itemize}
The range of the priors for the parameters is shown in  Table \ref{tab:param_prior_new}.

\begin{table*}
    \centering
    \caption{Prior ranges for expansion coefficients of model parameters. Part (a) shows polynomial parameters following $P(z) = P_0 + P_1 z + P_2 z^2$, and part (b) shows power-law parameters for $\beta(z) = \beta_0(1+z)^r + c$. All priors are uniform distributions.}
    \label{tab:param_prior_new}
    
    \vspace{0.3cm}
    \textbf{(a) Polynomial Parameters}
    \begin{tabular}{l|ccc|ccc|ccc|ccc}
    \toprule
    Parameter & \multicolumn{3}{c|}{$\alpha$} & \multicolumn{3}{c|}{$\log M_0$} & \multicolumn{3}{c|}{$\epsilon$} & \multicolumn{3}{c}{$\sigma_{\rm UV}$} \\
    \midrule
    Coefficient & $\alpha_0$ & $\alpha_1$ & $\alpha_2$ & $M_1$ & $M_2$ & $M_3$ & $\epsilon_0$ & $\epsilon_1$ & $\epsilon_2$ & $\sigma_0$ & $\sigma_1$ & $\sigma_2$ \\
    \midrule
    Prior & $[0.01,$ & $[-0.1,$ & $[-0.1,$ & $[8,$ & $[-1.0,$ & $[-0.01,$ & $[0.001,$ & $[-0.01,$ & $[-0.01,$ & $[-0.15,$ & $[-0.1,$ & $[-0.01,$ \\
    Range & $1.2]$ & $0.1]$ & $0.1]$ & $13]$ & $1.0]$ & $0.01]$ & $0.5]$ & $0.01]$ & $0.01]$ & $0.15]$ & $0.1]$ & $0.01]$ \\
    \bottomrule
    \end{tabular}
    
    \vspace{0.5cm}
    \textbf{(b) Power-law Parameter ($\beta$)}
    \begin{tabular}{l|ccc}
    \toprule
    Coefficient & $\beta_0$ & $r$ & $c$ \\
    \midrule
    Prior Range & $[0.01, 1.5]$ & $[0.01, 1.5]$ & $[0, 1.5]$ \\
    \bottomrule
    \end{tabular}
\end{table*}

\subsubsection{Observational Data}
Our analysis incorporates luminosity function measurements spanning $z = 4$ to $z = 16$, combining:
\begin{itemize}
    \item \textbf{HST data} ($z \leq 8$): From \citet{Vogelsberger2020}, including measurements from \citet{Bouwens_2021, Bouwens_2015, Oesch2018}
    \item \textbf{JWST data} ($z \leq 16$): From \citet{donnan_jwst_2024, McLeod2024, Harikane2024-spec, Harikane2024b-spec, Casey2024, Donnan2023,Harikane2023,Finkelstein2022}
\end{itemize}
These data sets include both spectroscopic and photometric redshift measurements, providing comprehensive coverage of the early universe's galaxy population.

\subsubsection{Model Comparison}
To determine which parameters require redshift evolution, we systematically compare models with different combinations of evolving parameters. We use information criteria to balance goodness-of-fit against model complexity \citep{Liddle_AIC}:

\begin{align}
    \text{AIC} &= -2\ln\mathcal{L}_{\rm max} + 2k \label{AIC} \\
    \text{BIC} &= -2\ln\mathcal{L}_{\rm max} + k\ln N \label{BIC} \\
    \text{DIC} &= D(\bar{\theta}) + 2p_{\rm D} \label{DIC}
\end{align}

where:
\begin{itemize}
    \item $k$ = number of free parameters
    \item $N$ = total number of data points
    \item $D(\theta) = -2\ln\mathcal{L}(\theta)$ is the deviance
    \item $p_{\rm D} = \overline{D(\theta)} - D(\bar{\theta})$ is the effective number of parameters
\end{itemize}

The model with the lowest information criterion value provides the best balance between fitting the data and avoiding over-parametrization. We consider differences $\Delta{\rm IC} > 5$ as providing strong evidence against more complex models. For assessing the goodness of fit, we use $\chi^2/\rm dof$. We calculate the effective number of degree of freedom using the formalism mentioned in Equation 29, \citep{dof}, which also takes account into the posterior covariance matrix.

This systematic approach allows us to identify which aspects of galaxy formation physics must evolve with redshift to explain the observed luminosity functions, providing physical insights into the changing conditions in the early universe.

\section{Results}\label{Results}

We begin our analysis with MCMC sampling for a baseline model where all key parameters ($\alpha$, $\beta$, $M_0$, $\sigma_{\rm UV}$, $\epsilon_0$) are free but held constant across all redshift bins. We adopt default values of $\alpha = 0.5$, $\beta = 0.6$, $M_0 = 10^{11}\,\mathrm{M}_{\odot}$, $\sigma_{\rm UV} = 0.5$, and $\epsilon_0 = 0.1$. While this constant-parameter model adequately fits the lower-redshift data points (Figure \ref{all free}), it fails to reproduce the high-redshift observations beyond $z>10$. The overall goodness-of-fit is poor, with $\chi^2/\mathrm{dof} = 1.94$ (see Table \ref{table1: information crit}). 
%\SA{Please inlcude p-value of this chi2 and dof.}

This systematic failure at high redshifts suggests that galaxy formation physics evolves significantly with cosmic time, requiring redshift-dependent parameters for a self-consistent framework. Consequently, we introduce redshift dependence in the star formation efficiency parameters ($\alpha$, $\beta$, $M_0$, $\epsilon$) and the UV scatter ($\sigma_{\rm UV}$), making the SFE a function of both $M_{\rm H}$ and $z$ within the double power-law formalism. We run MCMC chains for various models with different combinations of redshift-dependent parameters, fitting all redshift bins simultaneously.

\subsection{Best-Fitting Model and Joint Luminosity Function Analysis}\label{sec:best model}

As outlined in Section \ref{MCMC}, we explore two primary parametrization approaches: power-law and polynomial forms (up to quadratic terms). Initially, we focus on models with polynomial redshift dependence, where select parameters evolve with $z$ while others remain free but constant. We systematically test combinations starting with redshift-dependent $\beta$ and either $\epsilon_0$ or $\sigma_{\rm UV}$.

\begin{table}
\begin{center}
\caption{Model comparison results based on goodness-of-fit and information criteria. Parameters with redshift dependence are denoted as $f(z)$, while others are free but constant. $\beta(z)$ follows power-law parametrization unless otherwise indicated ($\beta(z)_{\rm poly}$), while other parameters use polynomial forms up to quadratic terms. Two models are shown where SFE$(z)$ is taken into account for UVLF fitting only at $z>10$. Model with $\alpha(z;\forall z>10)$ indicates a sharp transition from constant SFE to SFE$(z)$ at $z = 10$, where as model with $\alpha(z - 10);\forall~z>10$ indicates smooth transition at $z = 10$. Models are ranked by their overall performance, with the top model providing the best description of the data.}
\begin{tabular}{lccccc}
    \hline
Model & $\chi^2/\rm dof$ & $\Delta\rm AIC$ & $\Delta\rm BIC$ & $\Delta\rm DIC$ \\
\hline
$\alpha(z)$, $\beta_{\rm c}$, $\rm M_{0,c}$, $\sigma_{\rm UV,c}$, $\epsilon_{c}$ & 1.39 & 0.79 & 0 & 0  \\
\hline
$\alpha(z)$, $\beta_{\rm c}$, $\rm M_{0}(z)$, $\sigma_{\rm UV,c}$, $\epsilon_{c}$ & 1.34 & 0 & 4.63 & 1.16  \\
$\alpha(z)$, $\beta_{\rm c}$, $\rm M_{0,c}$, $\sigma_{\rm UV,c}$, $\epsilon(z)$ & 1.39 & 4.26 & 8.90 & 1.49  \\
$\alpha(z)$, $\beta(z)$, $\rm M_{0,c}$, $\sigma_{\rm UV}(z)$, $\epsilon_{c}$ & 1.40 & 6.29 & 16.36 & 1.54  \\
$\alpha(z)$, $\beta(z)_{\rm poly}$, $\rm M_{0,c}$, $\sigma_{\rm UV}(z)$, $\epsilon_{c}$ & 1.39 & 5.92 & 16.00 & 1.91  \\
$\alpha(z)$, $\beta(z)$, $\rm M_{0,c}$, $\sigma_{\rm UV,c}$, $\epsilon(z)$ & 1.41 & 8.34 & 18.42 & 2.92  \\
$\alpha_{c}$, $\beta_{c}$, $\rm M_{0}(z)$, $\sigma_{\rm UV,c}$, $\epsilon_{c}$ & 1.44 & 7.11 & 6.31 & 7.69  \\
$\alpha(z;\forall~z>10)$, $\beta_{\rm c}$, $\rm M_{0,c}$, $\sigma_{\rm UV,c}$, $\epsilon_{c}$ & 1.42 & 7.06 & 6.26 & 8.24  \\
$\alpha(z - 10);\forall~z>10$, $\beta_{\rm c}$, $\rm M_{0,c}$, $\sigma_{\rm UV,c}$, $\epsilon_{c}$ & 1.54 & 15.20 & 14.40 & 14.52  \\
$\alpha_{c}$, $\beta_{c}$, $\rm M_{0,c}$, $\sigma_{\rm UV}(z)$, $\epsilon_{c}$ & 1.66 & 7.11 & 6.31 & 29.60  \\
$\alpha_{c}$, $\beta_{c}$, $\rm M_{0,c}$, $\sigma_{\rm UV,c}$, $\epsilon_{c}(z)$ & 1.75 & 40.24 & 39.44 & 40.37  \\
$\alpha_{c}$, $\beta(z)$, $\rm M_{0,c}$, $\sigma_{\rm UV,c}$, $\epsilon_{c}$ & 1.93 & 58.65 & 57.85 & 56.85  \\
$\alpha_{c}$, $\beta_{c}$, $\rm M_{0,c}$, $\sigma_{\rm UV,c}$, $\epsilon_{c}$ & 1.90 & 54.62 & 48.38 & 54.34  \\
\hline
\end{tabular}
\label{table1: information crit}
\end{center}
\end{table}

\begin{table}
\begin{center}
\caption{Best-fit parameter values along with for our top two models ranked by information criteria. Best-fit values correspond to those minimizing $\chi^2$. The values indicated in square bracket is the median value of the parameter along with $\pm1\sigma$ bound wrt median. Model 1 has the fewest free parameters, while Model 2 includes additional redshift evolution for $M_0$.  Note that some of the posterior distribution of these parameters are non-gaussian, hence non uniform $\pm 1\sigma$ bound is present.
%\SA{Please include error on all these values.}
}
\label{tab:model_comparison}
\begin{tabular}{l c c}  
    \toprule
    & \textbf{Model 1} & \textbf{Model 2} \\
    & \textbf{$\alpha(z)$ only} & \textbf{$\alpha(z)$, $M_0(z)$} \\
    Parameter & \textbf{($z$-independent others)} & \textbf{($z$-independent others)} \\
    \midrule
    \multicolumn{3}{c}{\textit{Mass Scale Parameters}} \\
    \midrule
    $\log(M_0)$ & $11.30$ [$11.28^{+0.09}_{-0.08}$] & --- \\
    $M_1$ & --- & $10.80$ [$11.09^{+0.21}_{-0.16}$] \\
    $M_2$ & --- & $0.087$ [$0.053^{+0.033}_{-0.051}$] \\
    $M_3$ & --- & $0.004$ [$-0.001^{+0.006}_{-0.005}$] \\
    \midrule
    \multicolumn{3}{c}{\textit{Star Formation Efficiency Parameters}} \\
    \midrule
    $\epsilon_0$ & $0.27$ $[0.27^{+0.01}_{-0.01}]$ & $0.26$ $[0.27^{+0.01}_{-0.01}]$ \\
    $\alpha_0$ & $0.90$ [$0.94^{+0.14}_{-0.12}$] & $1.18$ [$1.00^{+0.13}_{-0.17}$] \\
    $\alpha_1$ & $-0.019$ [$-0.025^{+0.021}_{-0.022}$]  & $-0.091$ [$-0.043^{+0.041}_{-0.031}$]\\
    $\alpha_2$ & $-0.0015$ [$-0.0012^{+0.0010}_{-0.0010}$] & $0.0017$ [$-0.0002^{+0.0013}_{-0.0018}$]\\
    $\beta$ & $0.48$ [$0.47^{+0.07}_{-0.07}$] & $0.45$ [$0.47^{+0.07}_{-0.07}$] \\
    \midrule
    \multicolumn{3}{c}{\textit{Scatter Parameter}} \\
    \midrule
    $\sigma_{\rm UV}$ & $0.32$ [$0.27^{+0.12}_{-0.15}$] & $0.44$ [$0.33^{+0.10}_{-0.15}$] \\
    \bottomrule
\end{tabular}
\end{center}
\end{table}

Models incorporating redshift dependence in $\beta$, $\epsilon_0$, or $\sigma_{\rm UV}$ (with other parameters free but constant) provide reasonable fits to most data points, though the faint end could be better reproduced. This suggests the need for $\alpha$ evolution with redshift. When we include redshift-dependent $\alpha$ alongside the other evolving parameters (Figure \ref{alpha_beta_sigma_redshift}), the model successfully describes the data across all magnitude and redshift ranges. Similarly, maintaining the same redshift dependence while switching between evolving $\sigma_{\rm UV}$ and $\epsilon_0$ yields comparable quality fits.

We further explore models with power-law parametrization for $\beta$ ($\beta = \beta_0(1+z)^r + c$) while maintaining polynomial forms for other parameters. This approach provides unique insights into the necessity of $\beta$ evolution. Interestingly, $\beta$ shows minimal evolution with redshift, and the power-law feature is not prominent (Figure \ref{beta power}), yet the model still fits the data well. This weak evolution indicates that similar fitting quality can be achieved without requiring redshift dependence for $\beta$.

To identify the minimal parameter set, we systematically test the necessity of redshift dependence for each parameter. Removing the $z$-dependence of $\beta$ (making it free but constant across redshifts) maintains or even improves the $\chi^2/\mathrm{dof}$ in some cases. We also remove redshift dependence from $\epsilon_0$ and $\sigma_{\rm UV}$ in various combinations, seeking the minimal complexity model that provides good fits.

\begin{figure}
    \centering
    \includegraphics[width=0.7\linewidth]{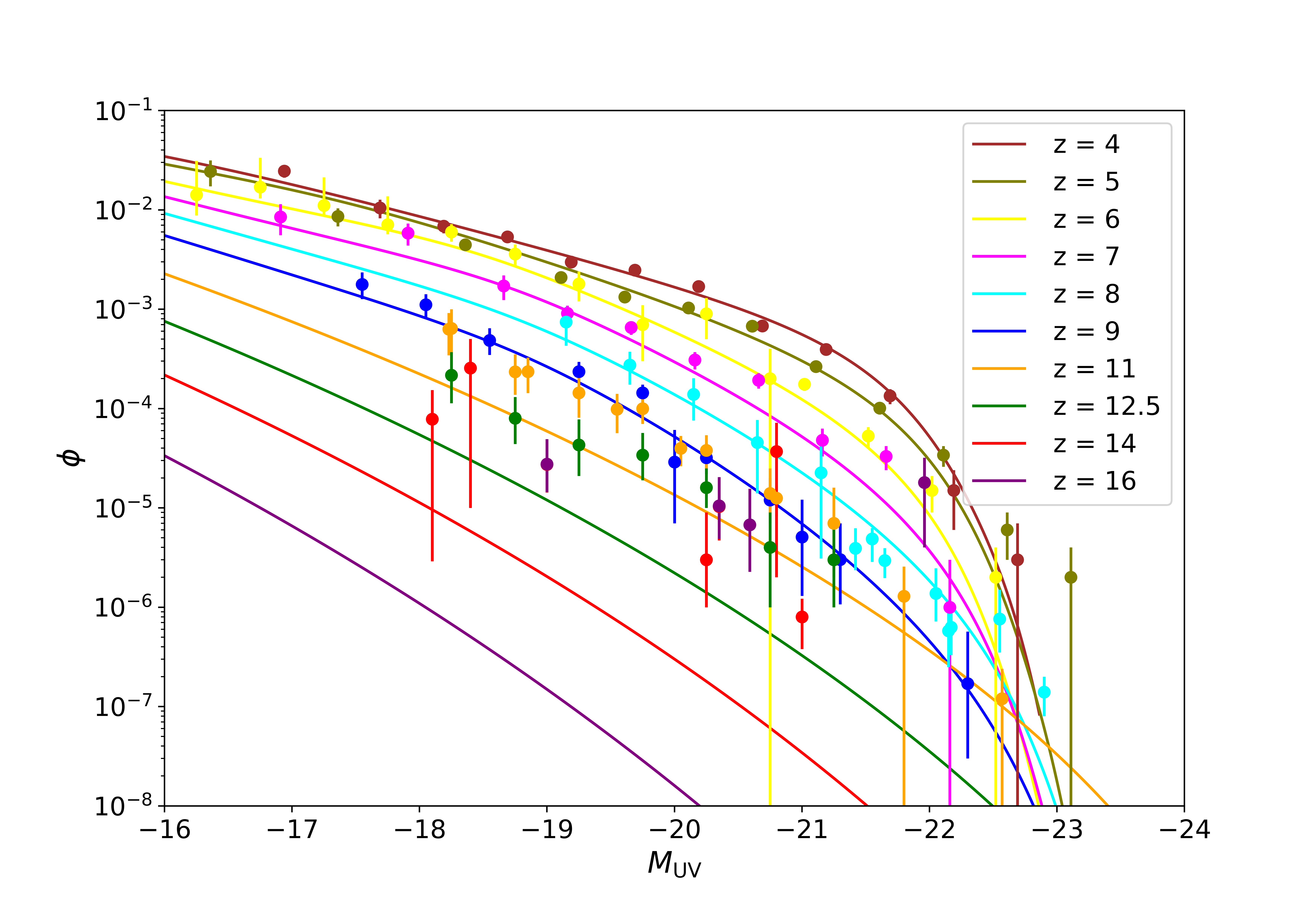}
    \caption{Joint fitting of LFs with model consisting of all free but constant parameter across the redshift range. The fitting shows the mismatch between model predicted and observational LFs at $z>10$ (see Section \ref{Results}).}
    \label{all free}
\end{figure}

\begin{figure}
    \centering
    \includegraphics[width=0.75\linewidth]{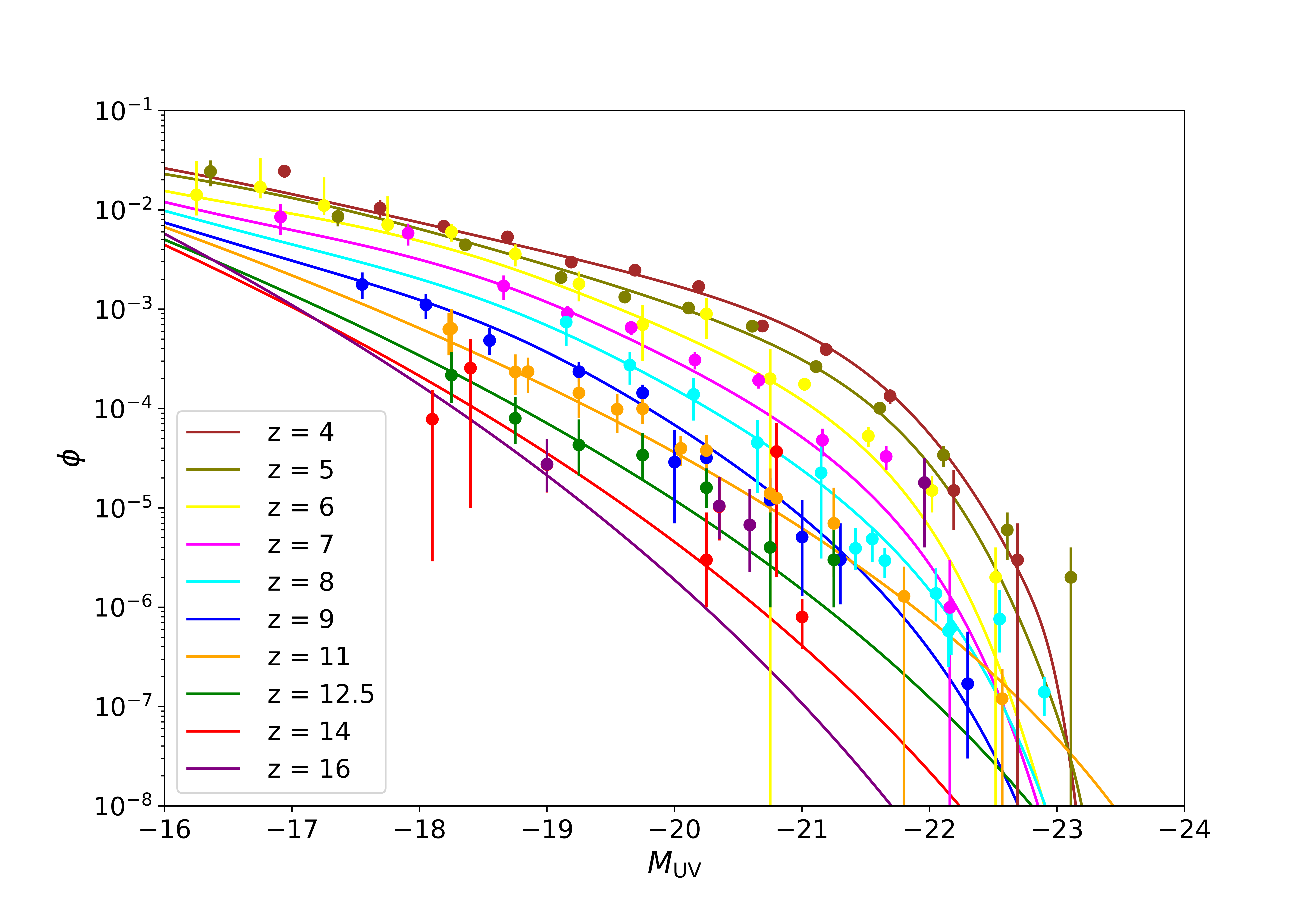}
    \caption{Best-fitting model (solid curves) with $\alpha$ having polynomial redshift evolution and other parameters free but constant. This requires only a small scatter $\sigma_{\rm UV}$ to obtain the final UVLF. The Y axis units are - number of galaxies/$\rm Mpc^{-3}/\rm mag^{-1}$. The data points are shown in circle along with error bars with different colors corresponding to different redshifts. This represents our optimal model requiring the minimum number of free parameters.}
    \label{alpha redshift LF}
\end{figure}

\begin{figure}
    \centering
    \includegraphics[width=0.65\linewidth]{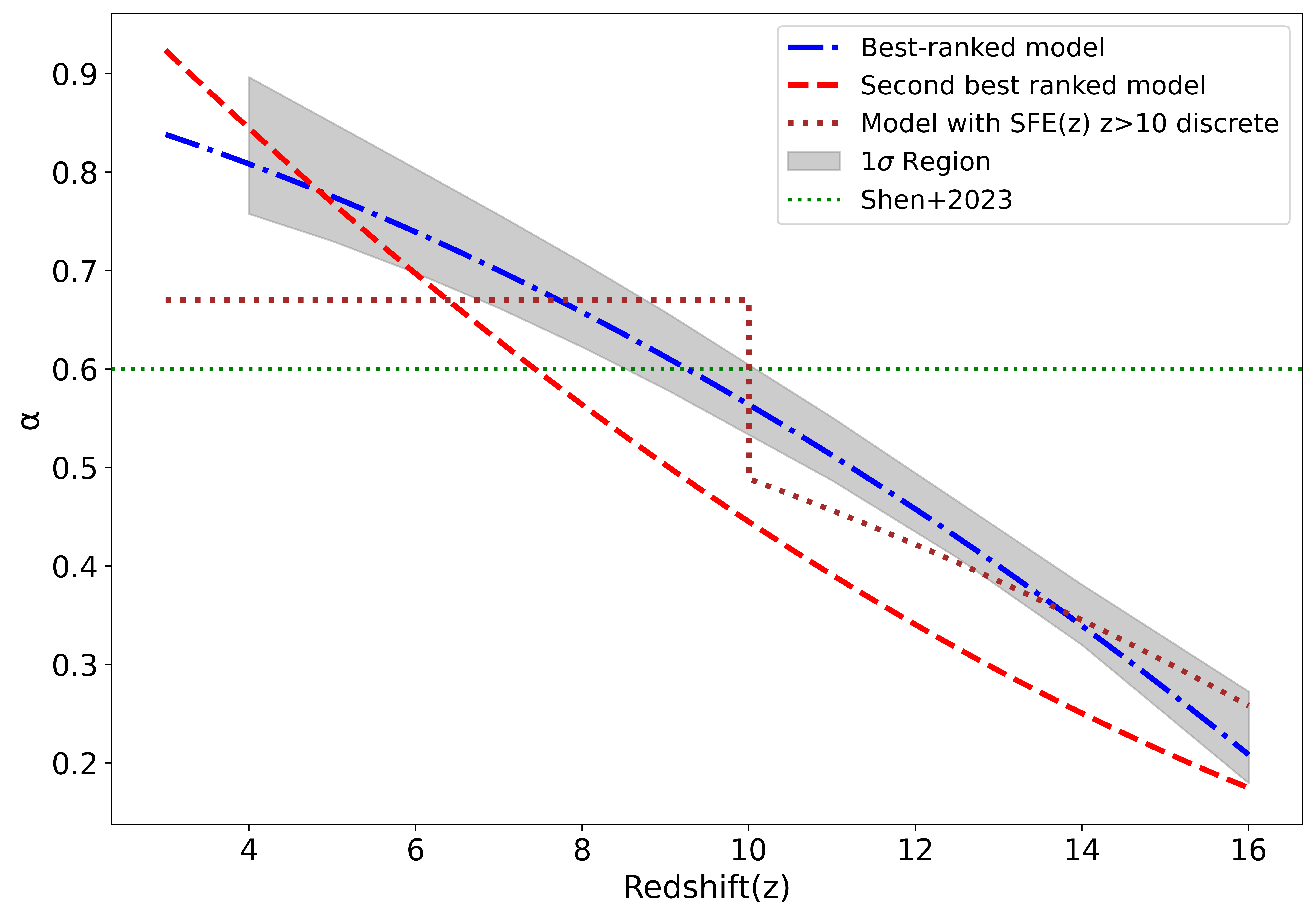}
    \caption{Evolution of $\alpha$ with redshift following polynomial form contrary to the constant value of $\alpha$ being taken in literature. $\alpha$ evolution for the model with SFE$(z;\forall z>10$) is shown in the plot too. The dashed blue line shows $\alpha$ evolution from best-ranked model, dashed red line shows the second best ranked model $\alpha$ evolution and the brown dotted line indicates $\alpha$ evolution for the SFE$(z;\forall z>10$) model. The horizontal dotted line indicates $\alpha = 0.6$ as taken in \citet{Shen2023}. The parameter shows strong evolution, demonstrating the importance of some degree of redshift dependence in the low-mass slope of the star formation efficiency.
    }
    \label{alpha poly}
\end{figure}

Our analysis reveals that redshift dependence is essential for $\alpha$ (polynomial function) to achieve good fits with the minimum number of free parameters (Figure \ref{alpha redshift LF}). Removing $z$-dependence from $\alpha$ significantly worsens the fit quality, with $\chi^2/\mathrm{dof}$ deteriorating from 1.39 to 1.90 (Table \ref{table1: information crit}) %\SA{could we also say p-values here}. 
This model shows strong $\alpha$ evolution with redshift (Figure \ref{alpha poly}). Further, since Fig \ref{all free} shows that the LF fitting remains good at $z\leq 9$ but fails at increasingly higher redshifts, we check a model with selective redshift dependence in SFE as mentioned in Section \ref{redshift evol}. Figure \ref{alpha poly} also demonstrates that the SFE model with $\alpha(z); \forall z>10$ shows a similar redshift evolution trend (as best ranked model) of $\alpha$ beyond $z = 10$, with constant $\alpha$ value before that being 0.67. This shows the necessity of having some level of redshift dependence in SFE for matching UVLFs.

In particular, our best-fit parameters differ significantly from previous literature values. We find best-fit values of $\sigma_{\rm UV} = 0.32$, $\log(M_0) = 11.3$, $\epsilon_0 = 0.27$, and $\beta = 0.48$, compared to the ad-hoc values $\epsilon_0 = 0.1$, $\alpha = 0.5$, $\beta = 0.6$ used in \citep{Shen2023, ShenEDE}. The modest value $\sigma_{\rm UV}$ demonstrates that high scatter is unnecessary: a balance between redshift-dependent SFE and moderate $\sigma_{\rm UV}$ can account for high-redshift observations, contrary to estimates in previous work \citep{Gelli2024, ShenEDE}.

From Table \ref{table1: information crit}, our second-rank model assumes that both $\alpha$ and $M_0$ are redshift dependent and perform very well in terms of information criteria and $\chi^2/\mathrm{dof}$. This model also yields a median $\sigma_{\rm UV} \approx 0.43$. Both top models are favored in the reduced $\chi^2$ and information criterion tests, with their best-fit parameter values detailed in Table \ref{tab:model_comparison}.

The model with both $\alpha(z)$ and $M_0(z)$ captures the evolution in both the low-mass end slope and the characteristic halo mass where SFE peaks. This model achieves $\chi^2/\mathrm{dof} = 1.34$, although at the cost of two additional free parameters, as reflected in the information criterion rankings. The resulting values of $\sigma_{\rm UV}$ and $\epsilon_0$ are 0.43 and 0.26, respectively, with both $\alpha$ and $M_0$ showing strong redshift evolution. The model with sharp/discrete SFE($z$) transition at $z =10$ performs well in overall quality of fitting as indicated by $\chi^2/\rm dof$ but is dis-favoured against other best-ranked model when $\Delta \rm IC$s are considered (Table \ref{table1: information crit}).

\begin{figure*}
    \centering
    \includegraphics[width=1.0\linewidth]{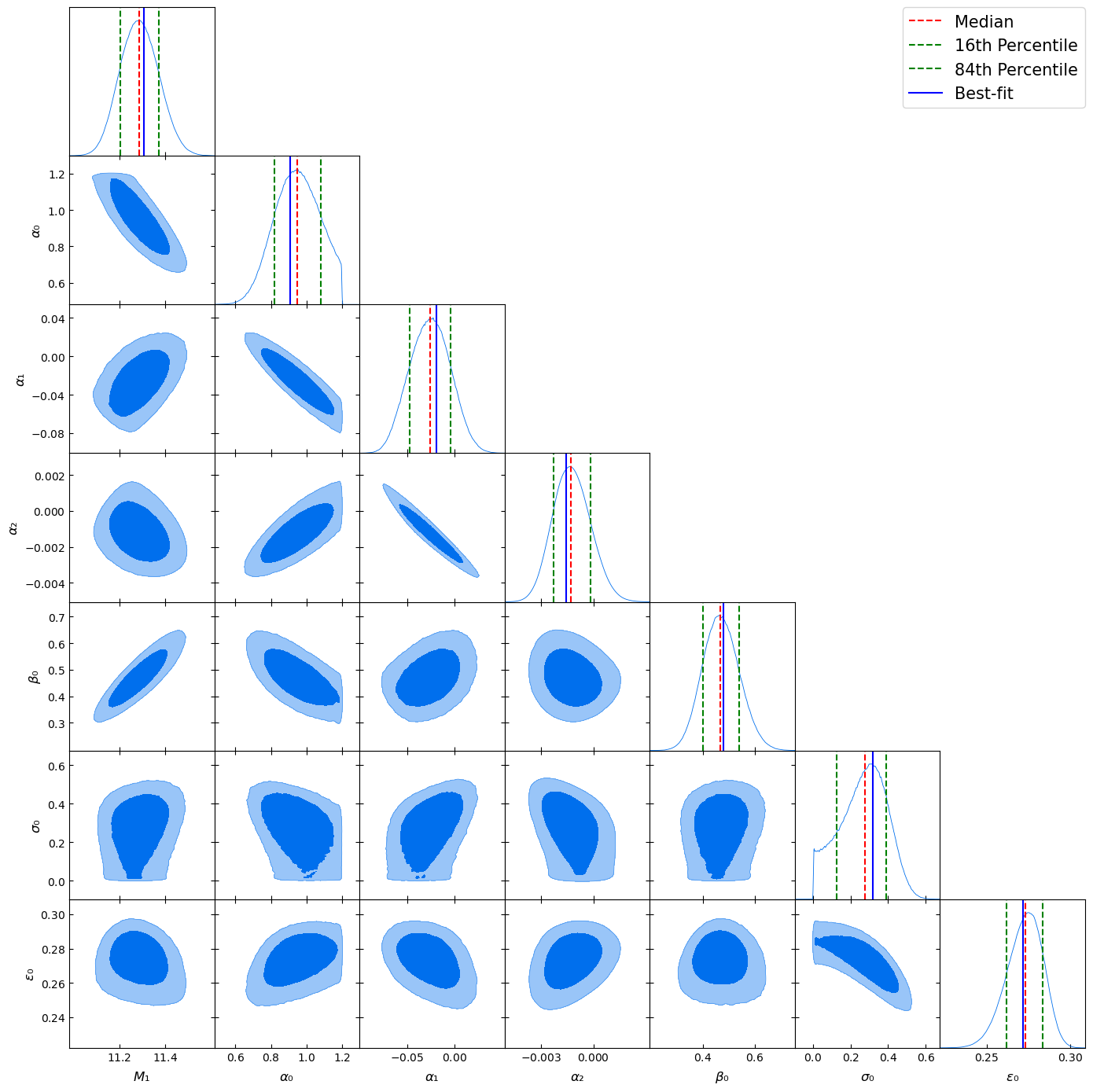}
    \caption{Parameter distribution and constraints for our best-fitting model with only $\alpha(z)$ evolving and other parameters free but constant across redshift. The tight contours demonstrate strong parameter constraints from the data.}
    \label{best model contour}
\end{figure*}

Considering both fit quality and parameter parsimony (Table \ref{table1: information crit}), we adopt the model with only $\alpha(z)$ as our best description. This choice is strongly favored by both BIC and DIC tests, while AIC and $\chi^2/\mathrm{dof}$ also indicate an excellent trade-off between model complexity and fitting quality. The model shows strong parameter constraints, as demonstrated in the contour plots (Figure \ref{best model contour}), with minimal sensitivity to the chosen prior ranges.

\subsection{Astrophysical Parameters and Relations}

Our model predictions enable derivation of various astrophysical observables, including star formation rates (SFR), specific star formation rates (sSFR), and stellar-to-halo mass relations (SHMR), which can be compared with available observational data.

\begin{figure}
    \centering
    \includegraphics[width=0.7\linewidth]{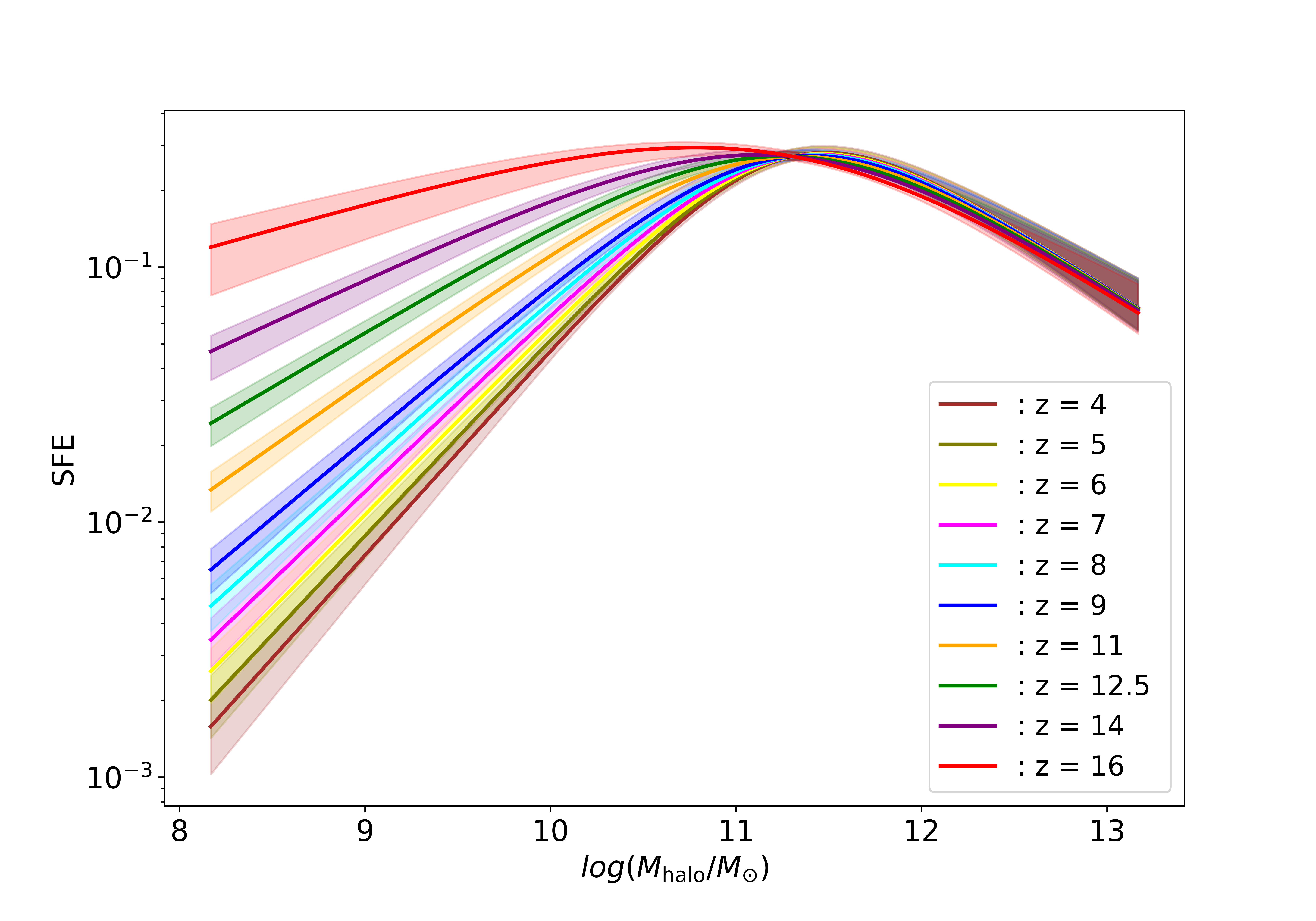}
    \caption{Star formation efficiency versus halo mass and its evolution across redshifts for the best ranked model. The SFE shows strong redshift evolution, particularly for lower-mass halos. The x axis halo mass is taken to be in log scale and in $M_{\odot}$ unit.}
    \label{SFE_fiducial model}
\end{figure}

The SFE evolution in our best model shows strong redshift dependence for halos up to $\sim 10^{11}\,\mathrm{M}_{\odot}$, driven by the evolving $\alpha$ parameter (Figure \ref{SFE_fiducial model}). This evolution appears consistent with theoretical predictions from \citet{Silk_2024}, which propose increasing SFE at high redshifts due to positive feedback from AGN outflows and subsequent "quenching" at lower redshifts below a transition redshift. Recent papers such as \citet{looser_notitle_2025} also discuss a local dip in SFH as ``mini quenching" phase.

For the second-ranked model with both $\alpha$ and $M_0$ redshift-dependent (Figure \ref{SFE_m1alpha}), we observe similar low-mass end evolution plus evolution in the halo mass achieving peak SFE. This scenario causes the peak SFE to shift with increasing redshift, consequently affecting the bright end of luminosity functions at higher redshifts.

\begin{figure}
    \centering
    \includegraphics[width=0.6\linewidth]{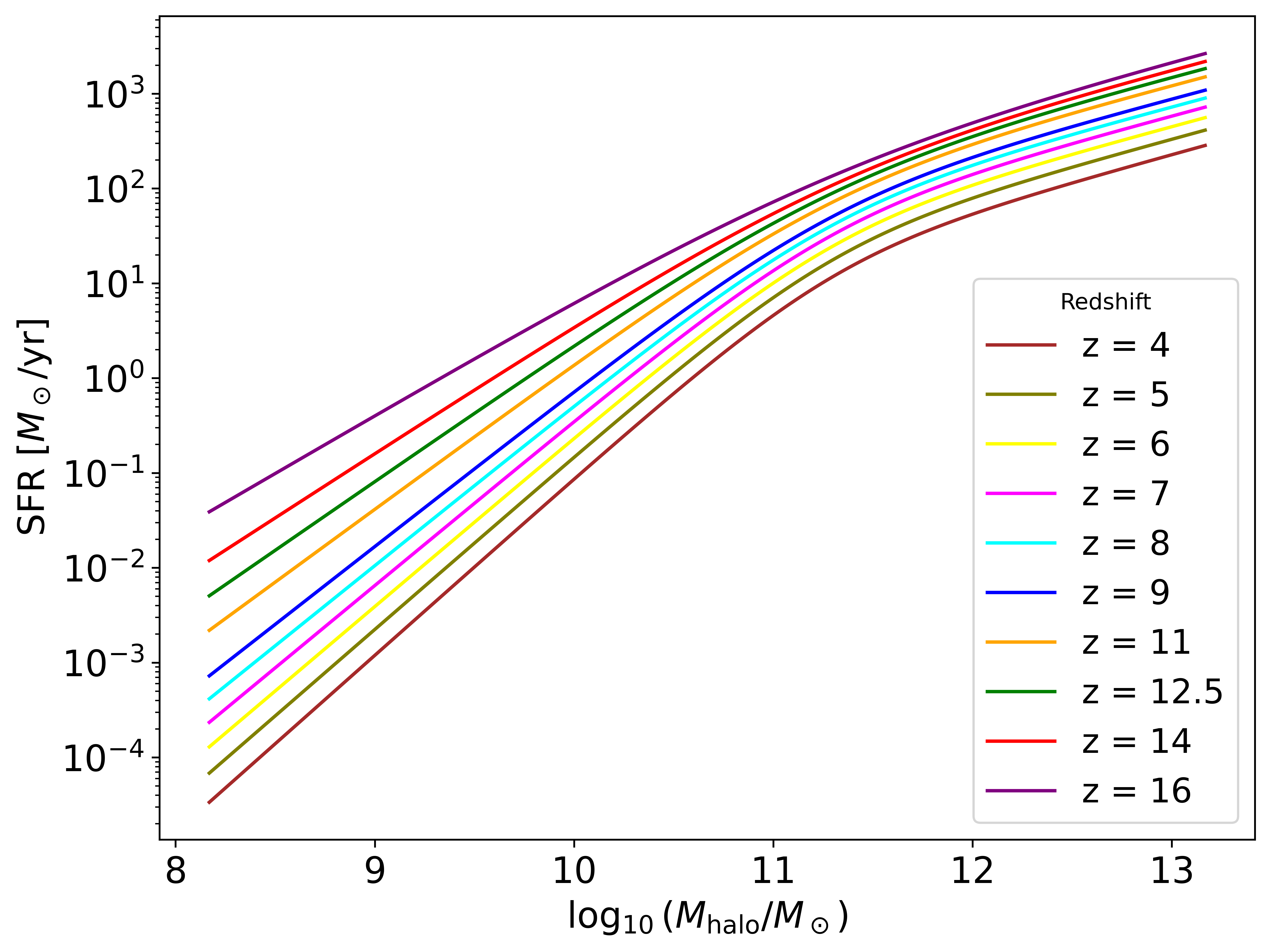}
    \caption{Star formation rate versus halo mass and its evolution across redshifts for the best ranked model. The SFR shows strong redshift evolution as expected from the SFE evolution. The x axis halo mass is taken to be in log scale and in $M_{\odot}$ unit.}
    \label{SFR_model}
\end{figure}

The SFR calculated (Figure \ref{SFR_model}) from the best-fit SFE denotes an increasing SFR as a function of redshift, in accordance with SFE evolution, which is particularly prominent at lower halo mass.
Regarding the role of stochasticity, \citet{shuntov2025sigma} report $\sigma_{\rm UV} \sim 0.6$ without redshift dependence (up to $z \leq 9$) and mildly evolving SFE. While their dataset is limited in redshift range, our best-fitting model demonstrates that a redshift-independent $\sigma_{\rm UV}$ coupled with more strongly evolving SFE can successfully account for UV luminosity functions across a wide redshift range. In Fig \ref{SFE_compare}, we also show a comparison between SFE($z$) - $M_{\rm H}$ relation from our best-ranked model and some of the parametrization considered in earlier works \citep[e.g][]{Harikane2022, Donnan2024, Tacchella2018, Shen2024}. Although the SFE-$M_{\rm H}$ relation is redshift independent, the comparison shows an increased SFE requirement in our model comparatively, especially at low halo mass end.

\begin{figure}
    \centering
    \includegraphics[width=0.7\linewidth]{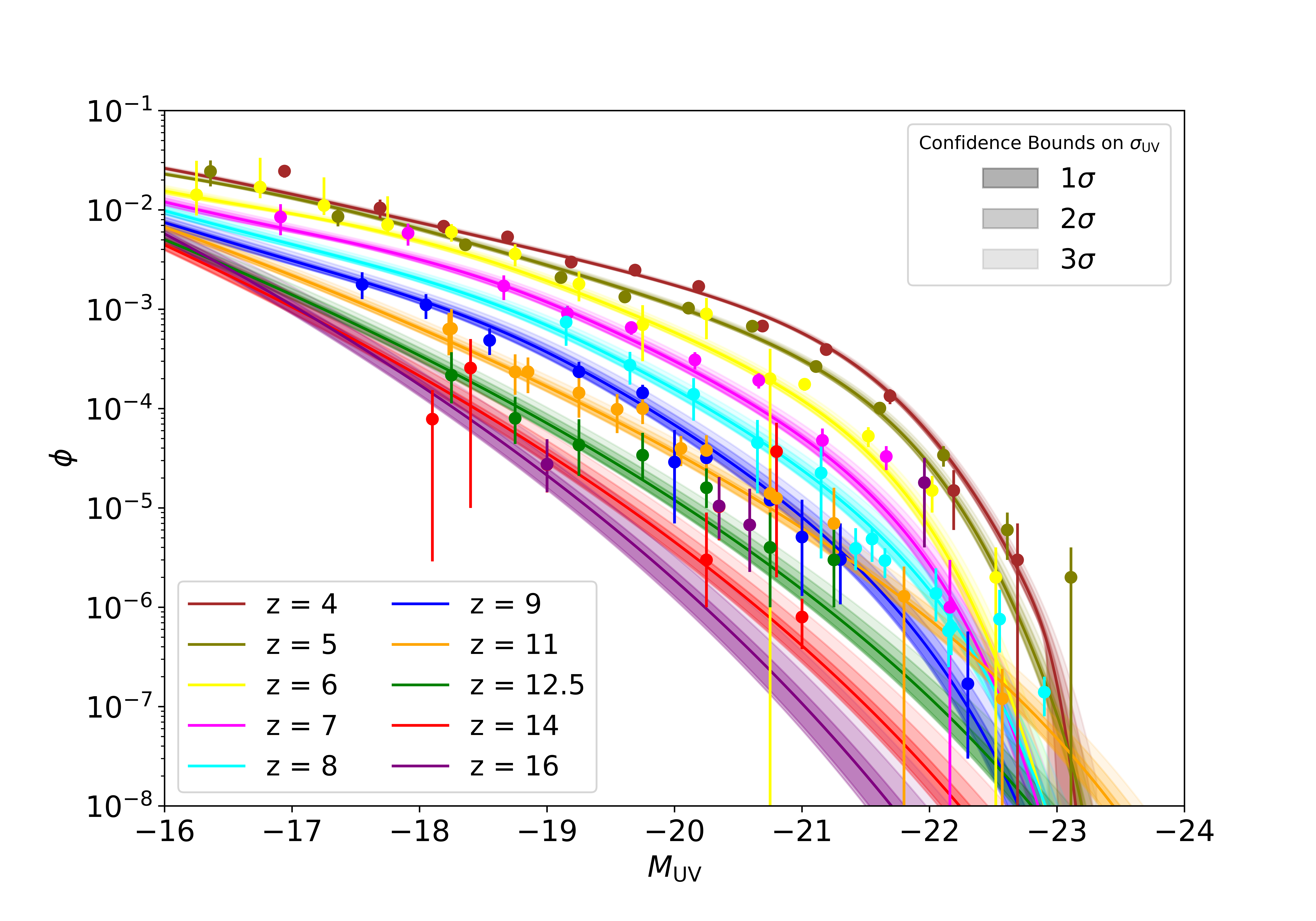}
    \caption{Sensitivity of UV luminosity functions to $\sigma_{\rm UV}$ variations in our best-fitting model. The Y axis units are - number of galaxies/$\rm Mpc^{-3}/\rm mag^{-1}$. The shaded regions show UVLFs corresponding to $\pm 1\sigma$, $2\sigma$, and $3\sigma$ confidence interval values of $\sigma_{\rm UV}$. The tight constraints indicate strong observational limits on star formation burstiness.}
    \label{sigma bound}
\end{figure}

\begin{figure}
    \centering
    \includegraphics[width=0.7\linewidth]{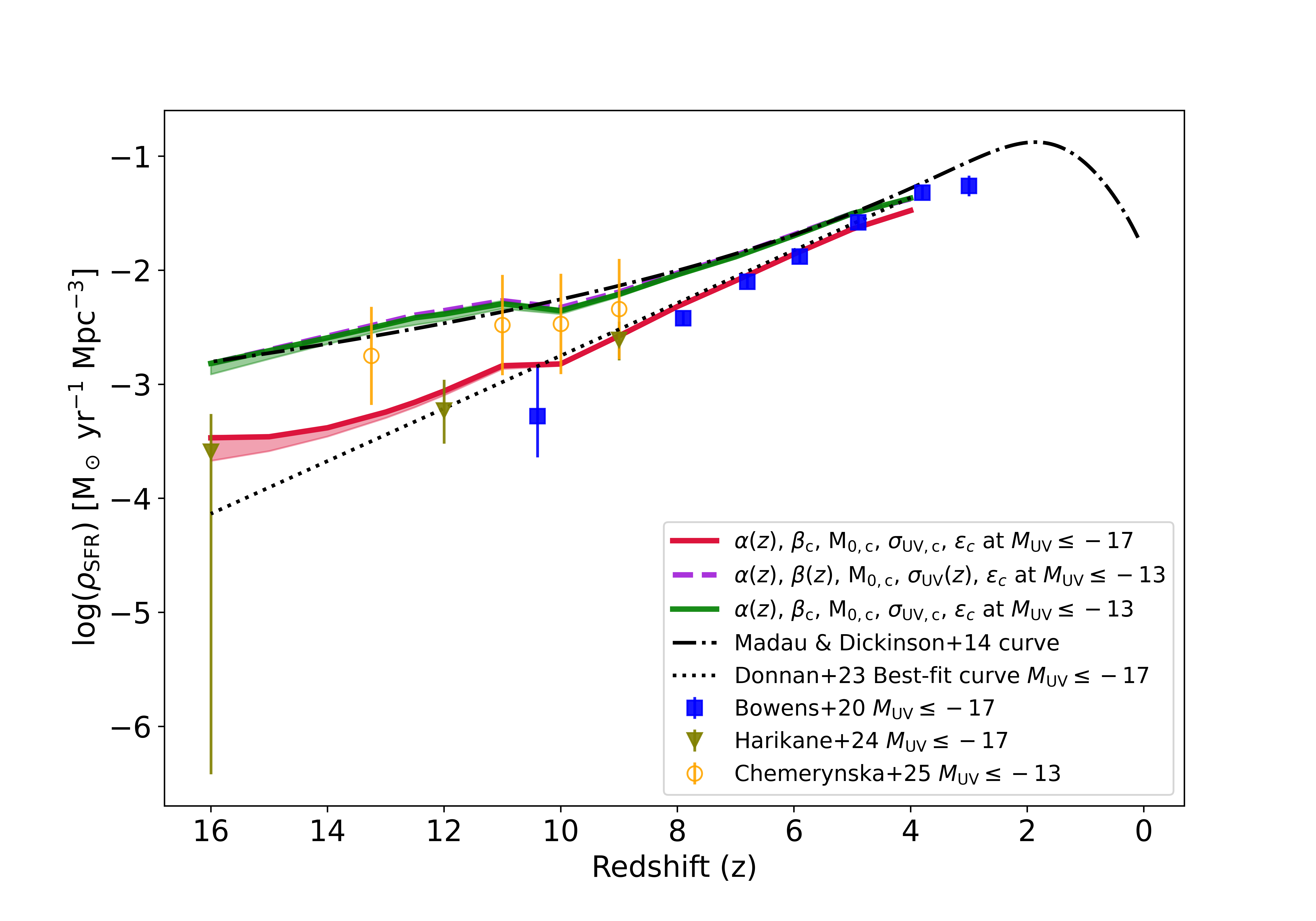}
    \caption{Star Formation Rate Density (SFRD) vs redshift as calculated from UVLFs of the best ranked model (i.e model with only $\alpha(z)$) for two different magnitude limits $M_{\rm UV}\leq -13$ and $M_{\rm UV}\leq -17$. SFRDs from another model including $\sigma_{\rm UV}(z)$ is also shown overlapping with best ranked model SFRD. %The shaded region indicates the $1~\sigma$ bound on the SFRDs from respective models. 
    Other data points and best fit curves shown are taken from \citet{Harikane2023,chemerynska_notitle_2025, Madau2014, Donnan2023, Bouwens2020}.}
    \label{SFRD fig}
\end{figure}

Figure \ref{sigma bound} demonstrates how UVLFs change with $\sigma_{\rm UV}$ variations, showing results for $1\sigma$, $2\sigma$, and $3\sigma$ confidence intervals. The UVLF variations remain well within observational error bars for different $\sigma_{\rm UV}$ values, indicating exceptionally strong constraints on this parameter and, by extension, on the level of stochasticity in star formation.

We also calculated cosmic Star formation Rate densities (SFRD) from our best-fit UVLFs, integrated upto different magnitude limits $M_{\rm UV} \leq - 13 $ and $M_{\rm UV}\leq -17$ along with another model including $\sigma_{\rm UV}$. As Figure \ref{SFRD fig} shows the SFRD evolution wrt redshift has shallow slope in both magnitude limits and SFRD estimates match between our best ranked model and the one having $\sigma_{\rm UV}(z)$ indicating that overall similar quality of fit and UVLF shape determine the SFRDs to be same. The SFRD evolution for $M_{\rm UV}\leq -13$ matches with predicted curve from \citet{Madau2014} and also within the error bars of data from recent most \citet{chemerynska_notitle_2025} which reports some faint end UVLF data points too. Further the calculated SFRD from our best ranked model shows very well match with earlier estimates from \citet{Bouwens2020, Harikane2023, Donnan2023} as well.

\subsection{Galaxy Bias Calculations}

Following Equation \ref{eq: bias}, we calculate the effective galaxy bias—a crucial parameter for large-scale structure studies—at our target redshifts. First, we determine the conditional probability distributions (Equation \ref{eq: PDF}), which identify the peak halo mass where the probability of finding a given $M_{\rm UV}$ is maximized.

\begin{figure*}
    \centering
    \includegraphics[width=0.7\linewidth]{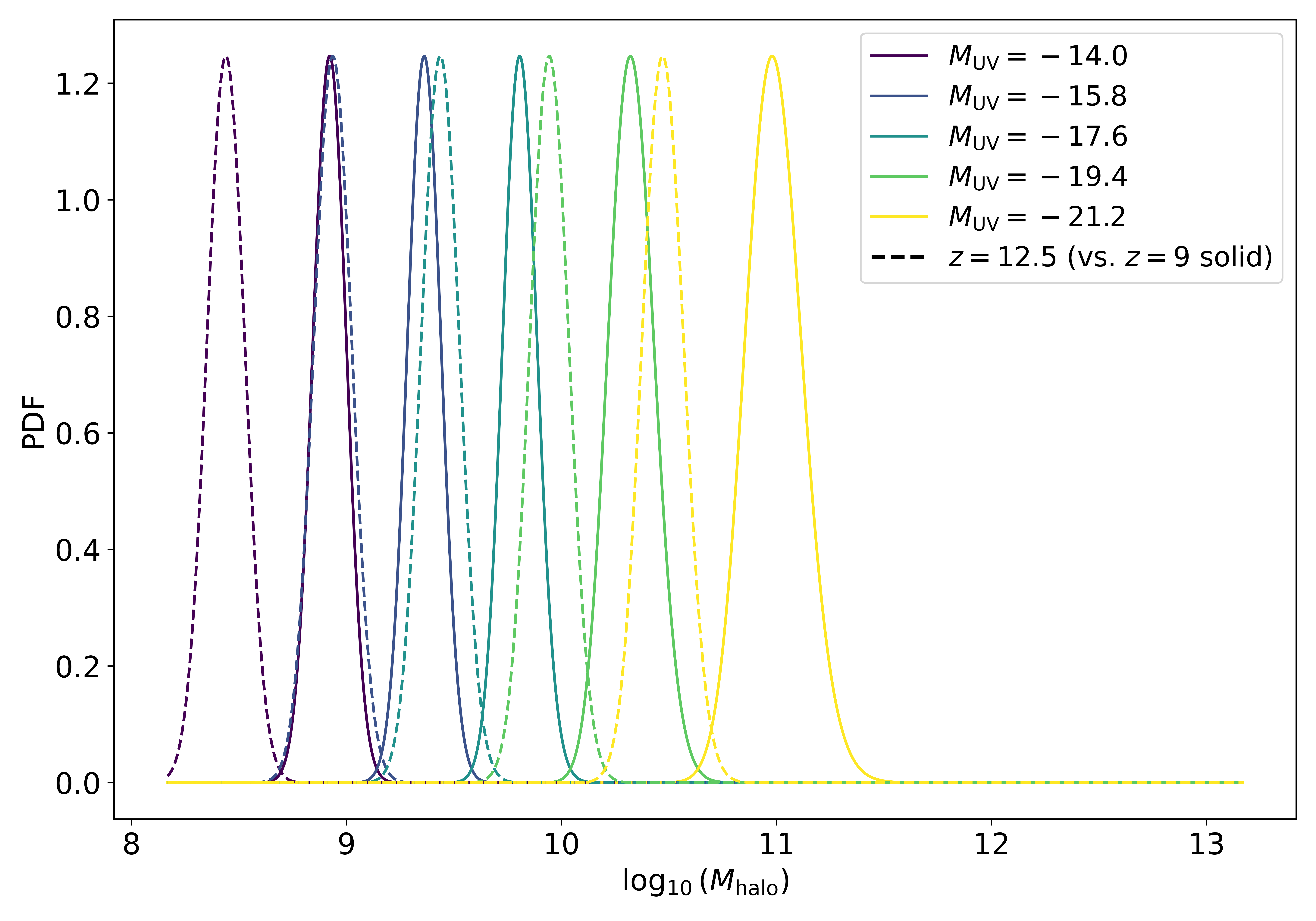}
    \caption{Conditional probability $P(M_{\rm UV}|M_{\rm H})$ versus $\log(M_{\rm H})$ for different UV magnitude limits at $z \sim 9$ (solid lines) and $z \sim 12.5$ (dashed lines). The distribution shifts toward lower halo masses for a given magnitude at higher redshifts. 
    }
    \label{fig:PDF}
\end{figure*}

Figure \ref{fig:PDF} illustrates these distributions for selected luminosity magnitudes at two different redshifts, highlighting the shift toward lower halo masses for a given magnitude at higher redshifts. Using these distributions and the halo bias $b(M_{\rm H})$, we calculate the effective bias $b_{\rm eff}$ for arrays of $M_{\rm UV}$ values.

\begin{figure*}
    \centering
    \includegraphics[width=0.6\linewidth]{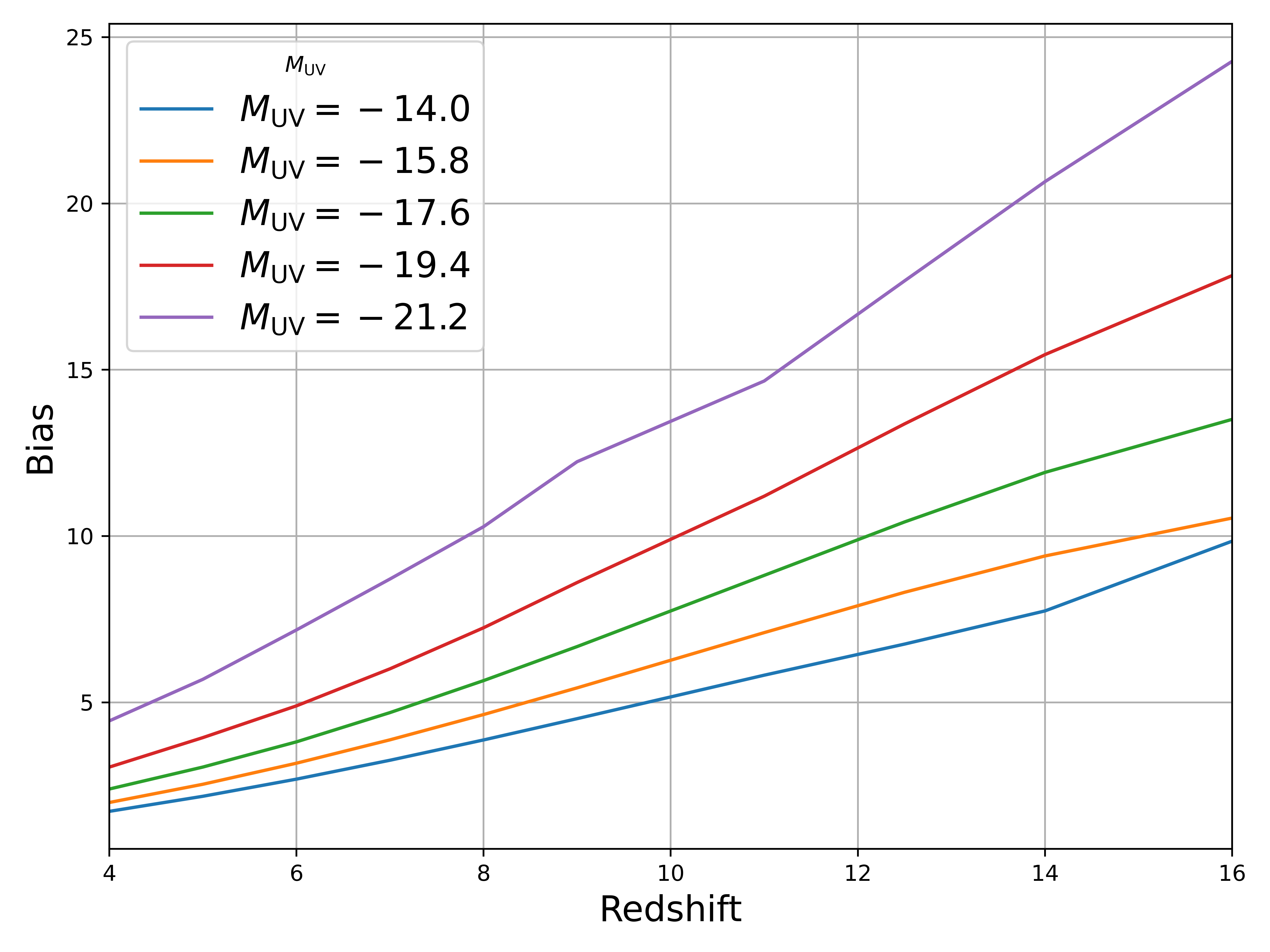}
    \caption{Galaxy bias from our best fit model as a function of redshift for different UV magnitudes, showing increasing trends with both redshift and brightness (more negative $M_{\rm UV}$).}
    \label{fig:bias_muv}
\end{figure*}

Figure \ref{fig:bias_muv} shows bias evolution with redshift for different $M_{\rm UV}$ values. As expected, bias increases with both redshift and brightness (more negative $M_{\rm UV}$), reflecting the fact that more massive halos host brighter galaxies at higher redshifts. We compute integrated bias for magnitude-limited samples with $M_{\rm UV} \leq -15.5$, $M_{\rm UV} \leq -19.1$, and $M_{\rm UV} \leq -19.8$.

\begin{figure*}
    \centering
    \includegraphics[width=0.6\linewidth]{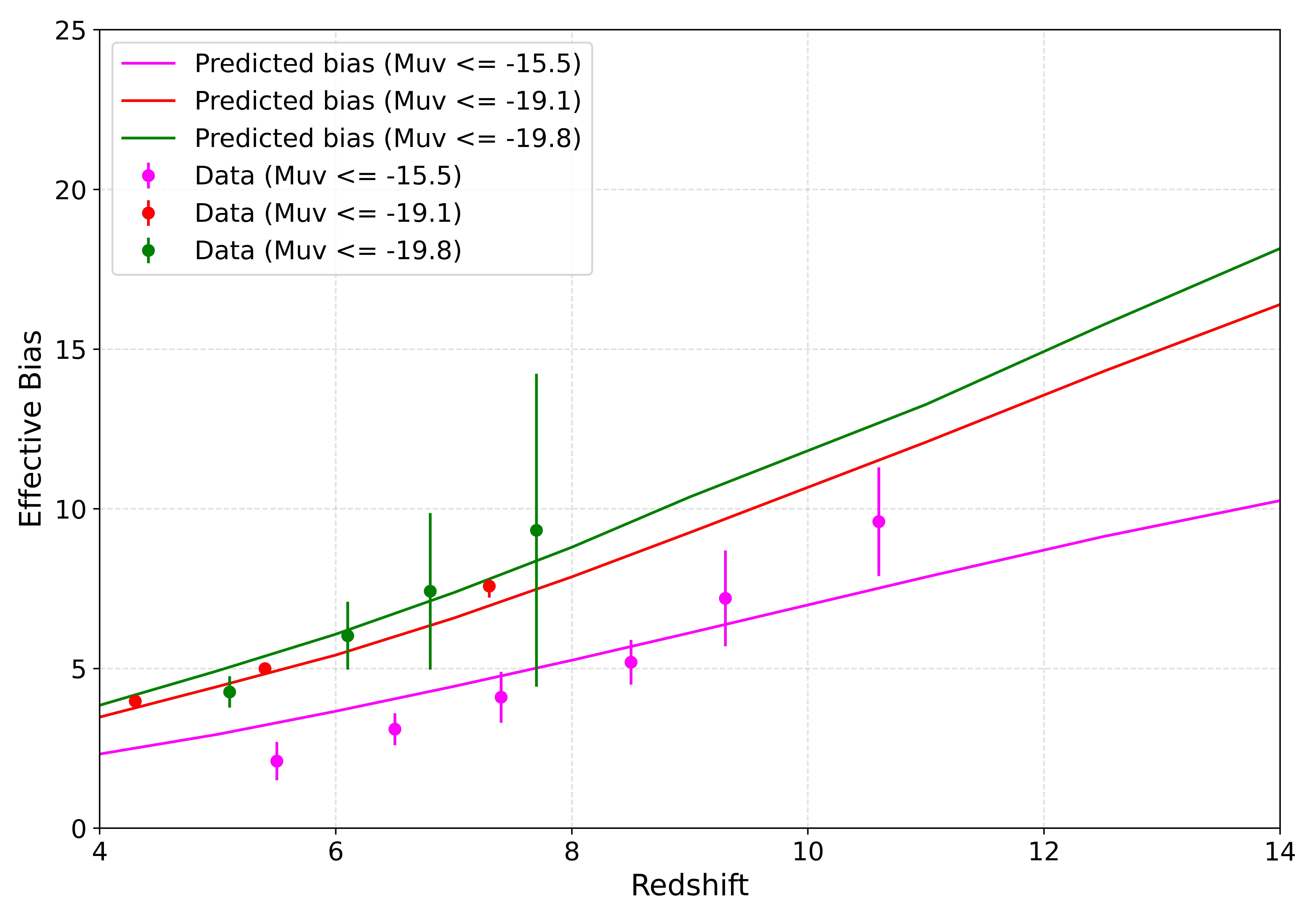}
    \caption{Integrated galaxy bias evolution with redshift for different magnitude limits ($M_{\rm UV} \leq -15.5$, $-19.1$, $-19.8$) compared with observational points. Data points are from \citet{dalmasso_notitle_2024}, \citet{shuntov_notitle_2025}, and \citet{dalmasso_galaxy_2024}, showing reasonable agreement with our predictions.}
    \label{fig:int_bias}
\end{figure*}

Figure \ref{fig:int_bias} compares our integrated bias calculations with observational data from \citet{dalmasso_notitle_2024, dalmasso_galaxy_2024, shuntov_notitle_2025}. The increasing bias trend with redshift agrees reasonably well with observations across all magnitude limits. We did not include bias in the optimisation process for this work but in future, this datapoints can be included in likelihood calculation and significantly improve the parameter estimates.

\section{Discussion}\label{sec: discussion}

\subsection{Model Predictions at the Highest Redshifts}

Having established our best-fitting model through comprehensive information criteria analysis (Section \ref{sec:best model} and Table \ref{table1: information crit}), we now examine its predictive power at even higher redshifts. Our optimal model, featuring redshift-evolving low-mass slope $\alpha(z)$ with the minimum number of free parameters, successfully describes observations up to $z = 16$. However, testing its extrapolation to $z > 17$ provides crucial insights into the model's validity and potential limitations.

We evaluate model predictions at $z = 17, 19, 25$ using preliminary data from recent JWST surveys. Several studies \citep{perez-gonzalez_notitle_2025, whitler2025, castellano_notitle_2025} have estimated luminosity functions from photometric observations at these extreme redshifts, though often only upper limits on number density are available, particularly for $z \gtrsim 20$.

\begin{figure*}
    \centering
    \includegraphics[width=1.1\linewidth]{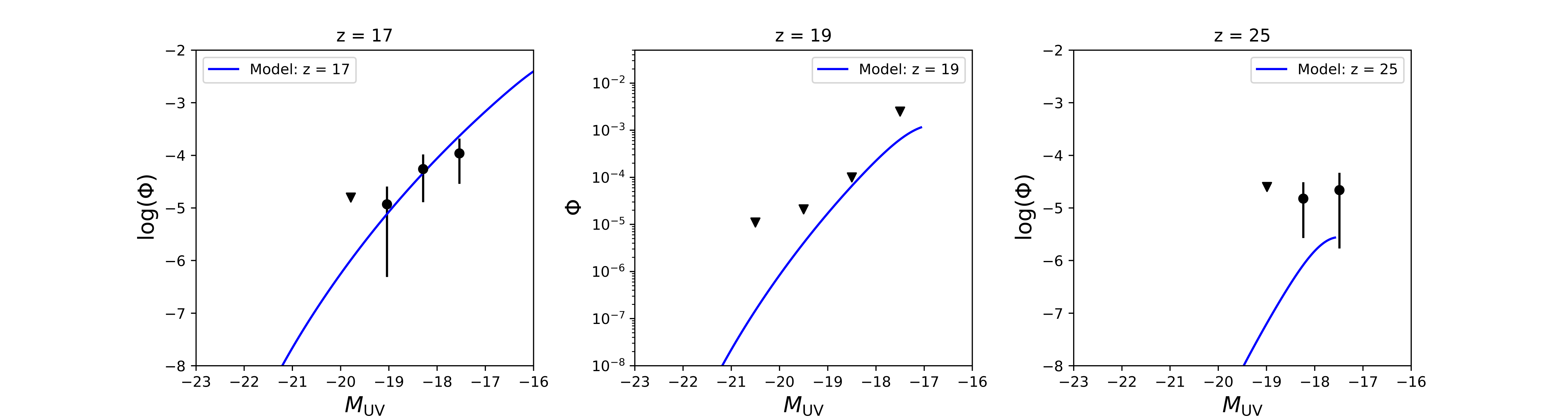}
    \caption{Model extrapolation for luminosity functions at $z = 17, 19, 25$. Data points at $z = 17, 25$ are from \citet{perez-gonzalez_notitle_2025}, while $z = 19$ observations are from \citet{whitler2025}. Black points with error bars represent detections, while triangles denote upper limits. The Y axis units are - number of galaxies/$\rm Mpc^{-3}/\rm mag^{-1}$.
    %\SA{I do not understand what you mean by y-axis scale match ... . Also, can you please add units on y-axis.}
    }
    \label{fig:LF_z17_19_25}
\end{figure*}

Figure \ref{fig:LF_z17_19_25} demonstrates our model's extrapolation capabilities up to $z \sim 25$. The predicted luminosity functions show excellent agreement with observations through $z \approx 20$, with our curves lying well within the observationally constrained upper limits. However, a significant discrepancy emerges at $z \sim 25$, where non-spectroscopic, and hence uncertain, observations seem to suggest higher galaxy number densities than our model predicts.

This breakdown at $z \gtrsim 20$ reveals important physical insights. While these extreme-redshift galaxies await spectroscopic confirmation, the preliminary upper limits pose intriguing challenges for semi-empirical models. Our best-fit framework achieves success through a delicate balance between elevated, redshift-dependent SFE and modest, constant scatter ($\sigma_{\rm UV}$). To match the $z > 20$ observations, if taken to be true without considering the spectroscopic determination, it would require even higher star formation efficiencies, possibly accompanied by increased scatter. However, with recent upcoming follow-up measurement several of these galaxies have shown to be low redshift interloper. This indicates that our model prediction cast further doubt on this extreme redshift luminosity function.
A particularly revealing feature appears at $z \sim 25$, where our predicted luminosity function begins at relatively bright magnitudes ($M_{\rm UV} \sim -17.5$, bottom panel of Figure \ref{fig:LF_z17_19_25}), while observations suggest substantial galaxy populations at these brightnesses. This limitation stems from the strong redshift evolution of the low-mass slope $\alpha$ (Figure \ref{alpha poly}). For $z \geq 20$, $\alpha$ approaches negative values, causing the star formation efficiency to lose its double power-law characteristics and exhibit high efficiency even in very low-mass halos. This eventually results in luminosity functions that are truncated at the bright end.

Addressing this limitation requires incorporating all available $z > 15$ data into future likelihood analyses and deriving updated parameter constraints through revised MCMC sampling. The systematic deviation at the highest redshifts suggests that either additional physical processes become important, or our parametric evolution forms require modification for extreme cosmic epochs. Alongside, it is to be noted that the sample size at such ultra high redshift is very sparse and often the constraints on these number densities can be unreliable.

The galaxy bias calculations provide an additional consistency check for our model framework. As demonstrated in Figure \ref{fig:int_bias}, our predicted effective bias values agree well with observations across different $M_{\rm UV}$ magnitude limits. Recent work by \citet{chakraborty_notitle_2025} emphasizes the importance of detailed star formation modeling and duty cycle arguments for matching galaxy bias observations, though their analysis extends to $z \sim 13$. This shows the non triviality of reproducing galaxy bias alongside obtaining galaxy LFs. In this context, our model successfully reproduces the LF evolution upto very high redshift and bias measurements with focus on important aspect of star formation and the burstiness providing confidence in the underlying framework. Future precision bias measurements at $z > 10$ will enable joint likelihood analyses combining luminosity function and clustering data for more robust parameter constraints.

\subsection{Star Formation Histories and Bursty Star Formation}

The role of stochastic star formation represents one of the most critical aspects for explaining high-redshift JWST luminosity function observations. The degree of burstiness is quantified by the scatter parameter $\sigma_{\rm UV}$, which has been the subject of considerable debate in recent literature.

Previous semi-empirical models \citep{Shen2023, ShenEDE} suggested the need for extremely high scatter ($\sigma_{\rm UV} > 2.0$ dex) to reproduce high-redshift observations—values exceeding upper limits from hydrodynamic simulations \citep{Feldmann2024, sun_notitle_2023, Katz2023}. Mass-dependent scatter models also fail to explain luminosity functions at $z \geq 13$ \citep{Gelli2024}. 

However, recent analyses point toward more moderate scatter requirements. \citet{Kravtsov2024} derive $\sigma_{\rm UV} \sim 1.2$ dex, while \citet{Pallottini2023} find $\sigma_{\rm UV} \sim 0.6$ dex from SERRA simulations. Observational analysis from \citet{shuntov2025sigma} report constant, modest scatter ($\sim 0.6$ dex) from FRESCO survey data limited to $z \leq 9$. Furthermore, very recently,  \citet{carvajal-bohorquez_notitle_2025} also find modest scatter of $\sigma_{\rm UV}\sim 0.5$ (with almost no redshift evolution) with detailed SED modeling, along with few SFE values $\leq 0.1$ for redshifts between 6 - 12. The few available SFE values around $\log(M_{\rm Halo})\leq 11.3$ seems to be consistent with our estimates at peak mass. However, a broad SFE - $M_{\rm Halo}$ relation and the slope of it at lower halo mass is required to be confirmed from wider range of observational data points in future, particularly from faint UVLF end. As shown in Fig \ref{SFE_compare}, the wide range of SFE values for lowest halo masses and hence unreliable, inferred from different UVLF fitting and semi-empirical models, indicates a necessity of obtaining full UVLF till the faintest end through observations. Since, the faint galaxies are connected to low mass halos through SFE slope $\alpha$, robust determination of faint end UVLF is very important for future surveys. For e.g a recent work by \citet{chemerynska_notitle_2025} provides some faint end LF data points, albeit photometrically and being limited by survey volume.

In order to understand the UVLF faint end predictions and in turn, the lowest halo mass end SFE values for which there is a wide uncertainty, one can make use of SFRD evolution upto different $M_{\rm UV}$ limits. As shown in Fig \ref{SFRD fig}, the SFRD evolution wrt redshift, upto $M_{\rm UV}\leq -13$ can provide a crucial hint when compared against recent most data from \citet{chemerynska_notitle_2025}. Given that \citet{chemerynska_notitle_2025} provides some UVLF faint end data and performs SFRD calculation using those, a match with our SFRD prediction integrating up to UVLF faint end (calculated from semi-empirical model) indicates a good inference of it and in turn, some bound on low mass end SFE. However, there are several other parameters such as IMF consideration, metallicity etc responsible for determining the UVLF faint end too, indicating a degeneracy among these. Among other possibilities, a difference due to underlying HMF can also be compensated with the SFE by the offset factor. For e.g, Fig \ref{HMF_compare} and Fig 5 of \citet{yung_characterising_2024} show an offset of factor 1.2 - 1.5 (or inverse) for different HMF calibrations on an average compared to the \citet{Tinker2008} HMF considered in our model, which would mean the SFE at respective halo mass should be enhanced/lessened by that factor.

Our theoretical analysis and comparison with UVLFs strongly favors the lower end of this range of $\sigma_{\rm UV}$. Both top-ranked models (Table \ref{table1: information crit}) prefer mass-independent $\sigma_{\rm UV} \sim 0.4-0.5$ dex to explain observed luminosity functions jointly through $z = 16$, with successful model predictions extending to $z \sim 20$. This reduced scatter requirement reflects a fundamental trade-off: lower stochasticity is compensated by higher star formation efficiency evolving with redshift (Figure \ref{SFE_fiducial model}) or atleast provides a new window into understanding the LF evolution vis-à-vis a possible small to modest amount of scatter required. This kind of trade-off with requirement of low stochasticity is also found in some recent results of simulations \citep[e.g, ][]{shen_thesan-zoom_2025}.

This trade-off carries important implications for AGN-galaxy connections. \citet{Silk_2024} propose that transitions between momentum-driven and energy-driven AGN outflows could drive SFE evolution—high efficiency at early times transitioning to quenching at lower redshifts. Along similar lines, \citet{gelli_notitle_2025} highlight the importance of studying quiescent galaxies, which dominate at fainter magnitudes and provide insights into bursty star formation histories. Some other works such as \citet{nikopoulos_notitle_2024, mauerhofer_notitle_2025} also highlights a increasing SFE model being crucial to explaining JWST observations, although within a different framework of IMF and increased dust enrichment.

Upcoming observations from COSMOS-WEB, FRESCO, GLIMPSE, BEACON, and other surveys will provide crucial tests of the modest-scatter, high-SFE scenario along with determining the full UVLF upto faintest of galaxies more robustly. Future analyses should also explore mass-dependent scatter within Bayesian frameworks to distinguish from simple constant-scatter models.

\subsection{Dust Attenuation Modeling}

\begin{figure*}
    \centering
    \includegraphics[width=0.7\linewidth]{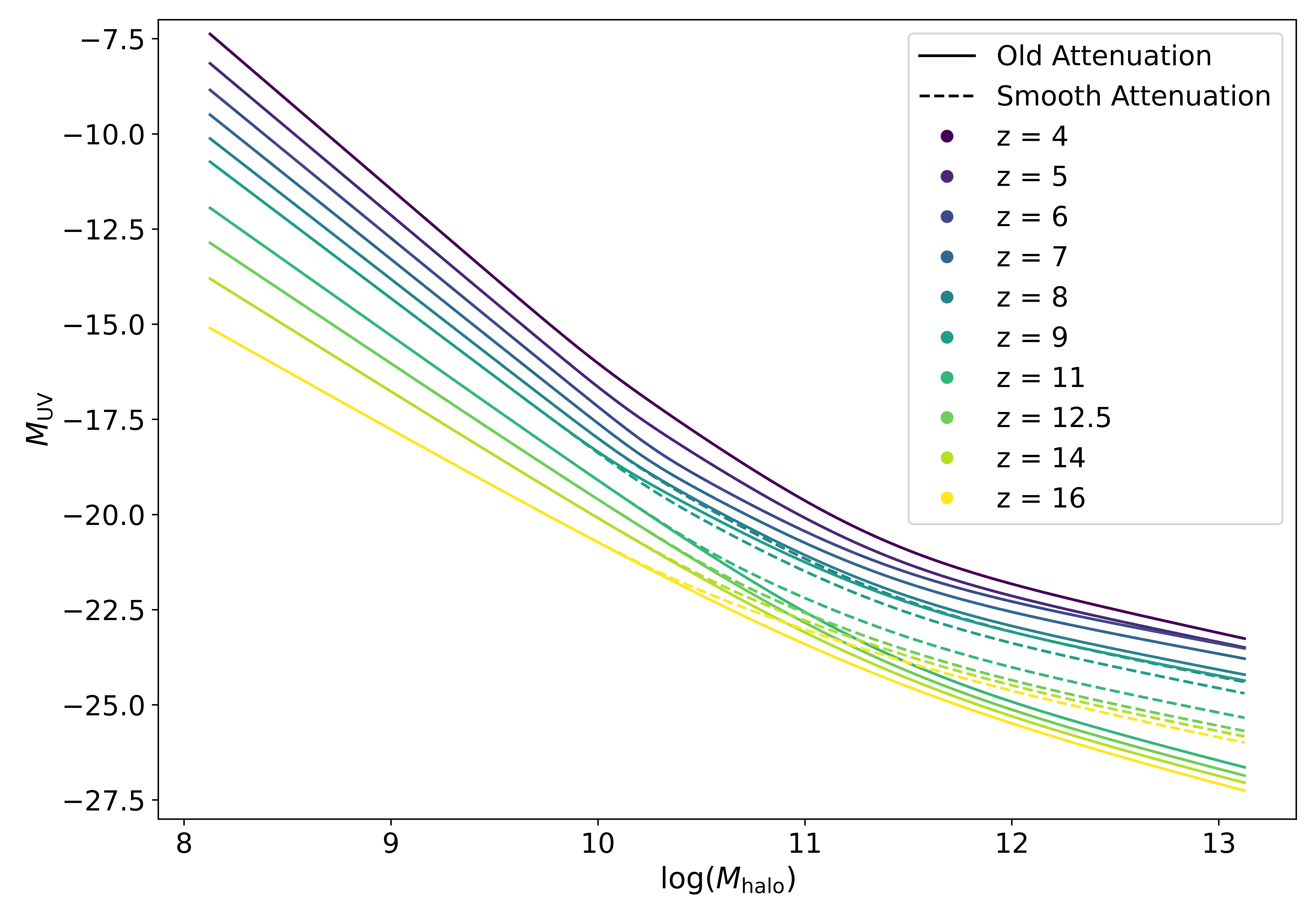}
    \caption{Comparison of dust attenuation models showing the $M_{\rm UV}-M_{\rm H}$ relation from the best model. The solid line represents our fiducial model with no attenuation for $z > 10$, creating a discontinuous slope change. The dashed line shows an example of smooth dust attenuation transition using interpolated values from \citet{Cullen2024}.}
    \label{fig:muv_mhalo_att}
\end{figure*}

Our dust attenuation prescription (Section \ref{sec: theory StandardLF}) assumes negligible extinction for $z > 10$, a reasonable approximation given the expected low dust content in early galaxies. However, this treatment introduces a discontinuous transition at $z = 10$ that creates artificial bimodality in the $M_{\rm UV}-M_{\rm H}$ relation (Figure \ref{fig:muv_mhalo_att}).

Recent spectroscopic observations by \citet{Cullen2024} provide UV slope measurements ($\beta_{\rm UV}$) extending to $z \sim 12$, indicating minimal but non-zero dust attenuation. To achieve smoother transitions, we interpolate these $\beta_{\rm UV}$ values, producing the gradual evolution shown by the dashed line in Figure \ref{fig:muv_mhalo_att}. This approach eliminates the artificial bimodality while maintaining the physical expectation of decreasing dust content at higher redshifts. Future refinements should incorporate smooth dust evolution models in luminosity function fitting. However, constraining dust properties at these extreme redshifts requires high-quality spectroscopic observations, as UV continua become increasingly blue with redshift \citep{morales_notitle_2023}. Some recent observations such as \citet{donnan_notitle_2025, tang_notitle_2025} find hints of reddening due to dust obscuration at early redshift  $z\geq 9$. More  observations are needed to conclusively arrive at the level of dust attenuation at increasingly higher redshifts along with their potential astrophysical implications. 

Further, one can connect the extent of dust obscuration with consideration other astrophysical phenomena such as SNe in the early galaxies and the impact of metallicities. \citet{mckinney_notitle_2025} show that a detailed consideration of dust attenuation laws in the context of SNe dust can impact the physical properties of the galaxies observed at $z \sim 6 - 12$. Especially if we consider detailed modelling, one can expect more dust formation during earlier times as more massive, metal poor stars with short lifetimes would undergo SNe phases. This would also self-consistently describe the evolution of metallicity (or lack of it) across higher redshifts, especially in the context of recent observations, e.g. \citet{morishita_notitle_2025},  highlighting the presence of a relatively metal-free environment even at comparatively later redshifts ($z\sim 5$). This might change the dust attenuation formalism   considered in recent works.

\subsection{Summary and Implications}

Our analysis yields several key insights for early galaxy formation:

\begin{itemize}
    \item \textbf{Reduced scatter requirements}: We find that modest, redshift-independent scatter ($\sigma_{\rm UV} \sim 0.4 - 0.5$ dex) suffices to explain observations through $z \sim 19$. This significantly reduces the need for extreme burstiness ($\gtrsim 1.3$ dex) suggested by earlier studies for $z > 13$, resolving tensions between theoretical predictions and JWST observations within standard $\Lambda$CDM cosmology.
    
    \item \textbf{Evolving star formation efficiency}: Our best model requires the SFE to increase with redshift, particularly in low-mass halos (driven by $\alpha(z)$ evolution). This supports theoretical scenarios of enhanced star formation at cosmic dawn, enabling rapid early galaxy growth as indicated by sSFR evolution. Maximum SFE reaches $\sim 0.2$ in our framework.
    
    \item \textbf{SFE-scatter trade-off}: The balance between elevated SFE and moderate scatter successfully reproduces luminosity functions without invoking extreme parameter values. Derived quantities (e.g, SFR) show excellent agreement with observational constraints, providing confidence in the physical framework.
    
    \item \textbf{Dust modeling limitations}: Our dust attenuation prescription creates artificial discontinuities in the $M_{\rm UV}-M_{\rm H}$ relation at $z \sim 10$. Future work requires more sophisticated dust evolution models based on UV continuum observations at extreme redshifts.
    
    \item \textbf{Predictive limits}: Model extrapolation succeeds through $z \sim 19$ but fails to reproduce preliminary constraints at $z \sim 25$, should such redshifts eventually be confirmed. Incorporating these extreme-redshift data into likelihood analyses may require higher SFE values or modified parametric evolution forms.
    
    \item \textbf{Clustering and SFRD consistency}: Our model predictions agree well with available galaxy bias measurements (Figure \ref{fig:int_bias}). Future precision clustering observations across wider redshift ranges can be combined with luminosity function data in joint Bayesian analyses to strengthen parameter constraints. SFRD calculations from our best ranked model matches quite well with other SFRD values from literature, especially with faint end UVLF considered in some recent works.
\end{itemize}

Our results demonstrate that JWST's early galaxy observations can be understood within the standard cosmological framework through physically motivated evolution of the star formation efficiency, without requiring extreme stochasticity or exotic physics. However, one important aim should be to obtain more observational data to actually compare with the star formation history, and most notably with our  proposed evolution of the SFE - $M_{\rm Halo}$ relation. This could further pin down the characteristics of star formation history across cosmic  redshift along with more robust constraints on other diagnostics such as the stellar-to-halo mass relation (SHMR) etc. as a function of $z$. Also, the emerging tensions at $z \gtrsim 20$ highlight the need for continued model refinement as observations push towards cosmic dawn.

\begin{acknowledgments}
We thank anonymous referee for insightful comments and suggestions which led to improvement of the manuscript. We thank the authors of different papers for providing with the necessary observational data related to UVLF and others. We also thank Xuejian Shen for making the code of the semi-empirical model, public and modifiable and Navin Chaurasiya for the discussions and code related to statistical calculations. Further, we thank Stuart Wyithe, Pierluigi Rinaldi, Tirthankar Roy Choudhury for insightful comments which helped improve the manuscript. This work is based on a masters' thesis by the corresponding author AK, presented at IISER Berhampur.
\end{acknowledgments}

\section{Data Availability}
The code for MCMC sampling including the Information Criteria rankings, $\chi^{2}$ calculation, plotting is available in the public \texttt{``EASYmcmc" }\href{https://gitlab.com/shadaba/easymcmc/-/tree/master/examples/UVLF?ref_type=heads}{repository}.

\appendix

\section{Other Models}\label{appendix}
Here we show the plots of joint LF fitting for other models except the best ranked models. These models provided the necessary intuition to understand the improvement needed further for getting the best one. We also show key astrophysical parameter : SFE evolution wrt $z$ for our next best ranked model.

\renewcommand{\thefigure}{A\arabic{figure}}
\setcounter{figure}{0}
\begin{figure}
    \centering
    \includegraphics[width=0.7\linewidth]{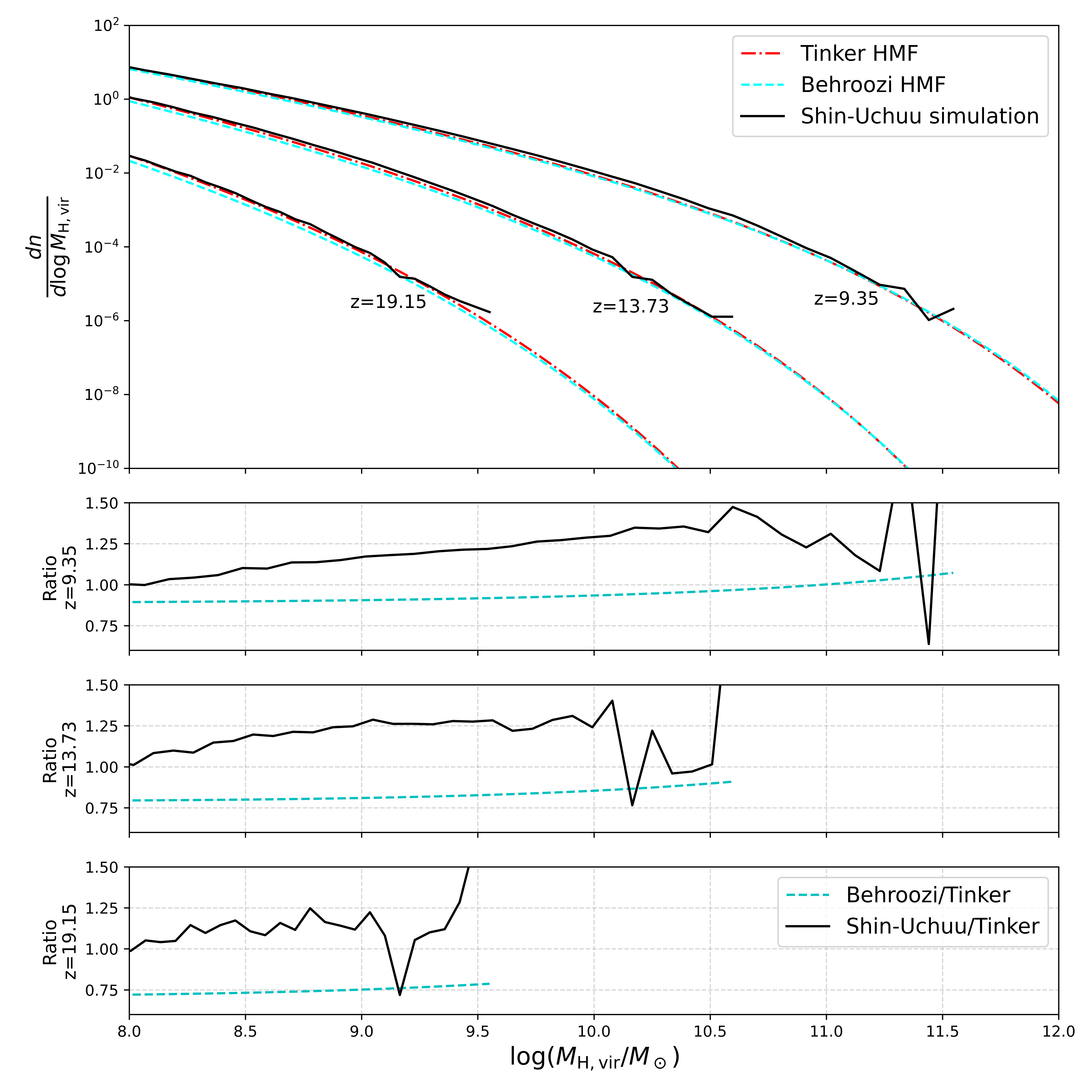}
    \caption{Comparison between different calibration of HMF and inference from Shin-Uchuu simulation. The main plot shows comparison of HMF at different high redshifts, in case of \citet{Tinker2008, Behroozi2013} and results of our fittings from Shin-Uchuu simulation halo catalogue \citet{Ishiyama2021}. The bottom horizontal plots indicate the ratio between different HMFs wrt \citet{Tinker2008} considered as baseline. Both of the HMFs show only a offset of factor 1.2 - 1.5 when compared with \citet{Tinker2008}.}
    \label{HMF_compare}
\end{figure}

\begin{figure}
    \centering
    \includegraphics[width=0.7\linewidth]{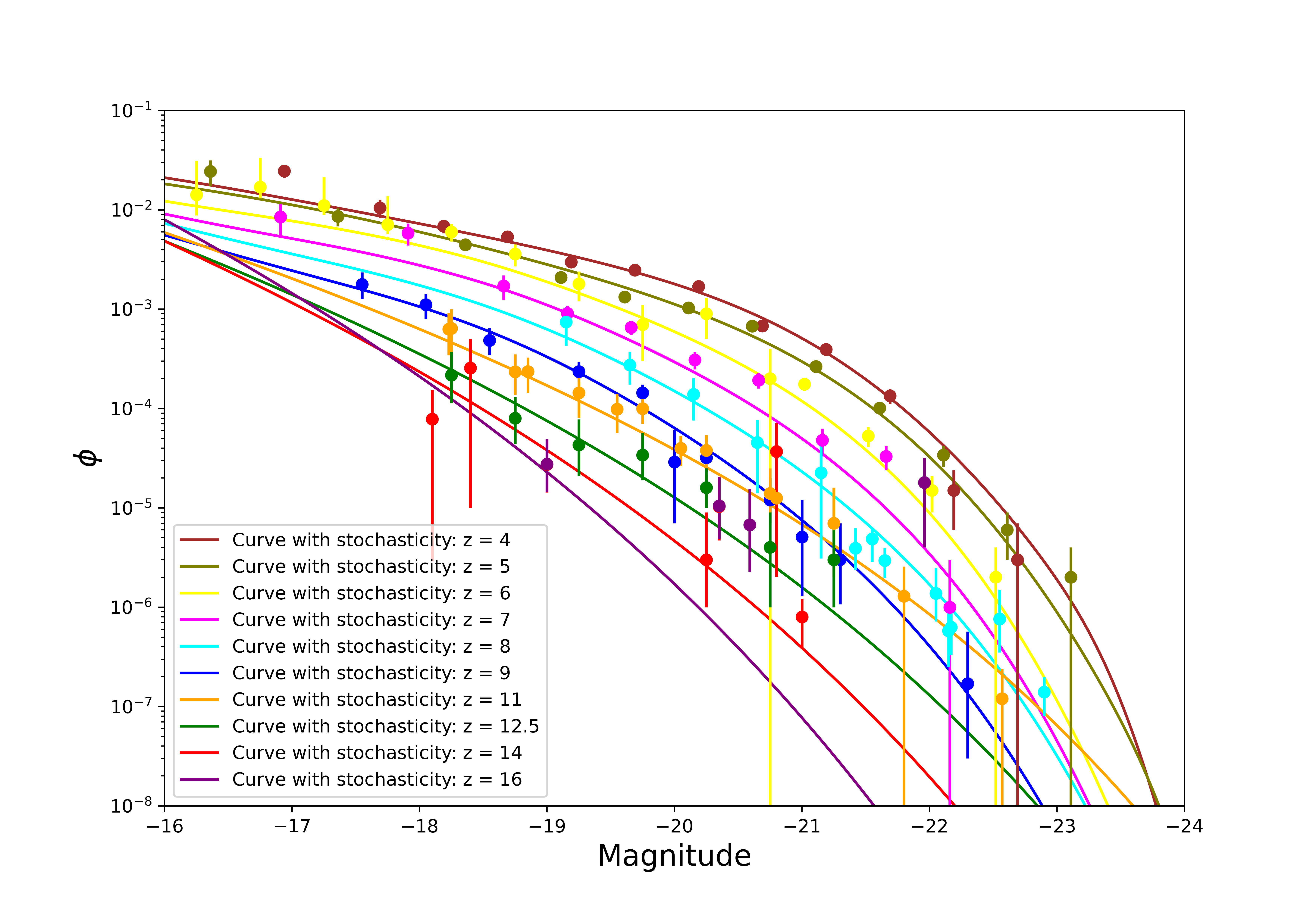}
    \caption{Model with $\beta$, $\alpha$, $\sigma_{\rm UV}$ redshift dependence and other parameters ($\epsilon_{0}$) being free but constant across redshifts. $\beta$ is taken to be in polynomial parametrization.}
    \label{alpha_beta_sigma_redshift}
\end{figure}

\begin{figure}
    \centering
    \includegraphics[width=0.7\linewidth]{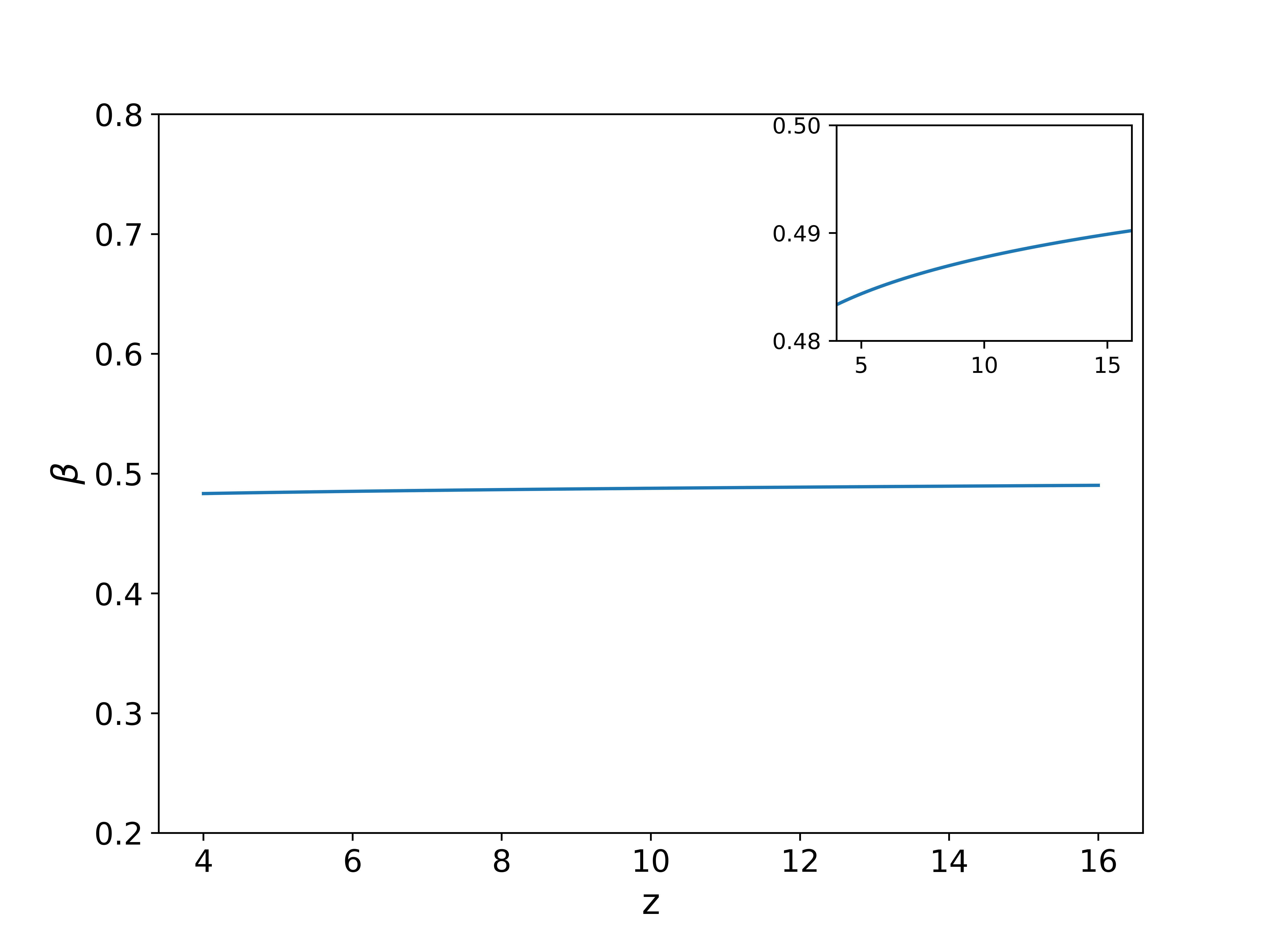}
    \caption{Evolution of $\beta$ wrt $z$ having power law evolution, which is almost constant across redshifts (Section \ref{Results}). The zoomed -in inset figure shows little evolution of the parameter value.}
    \label{beta power}
\end{figure}

\begin{figure}
    \centering
    \includegraphics[width=0.7\linewidth]{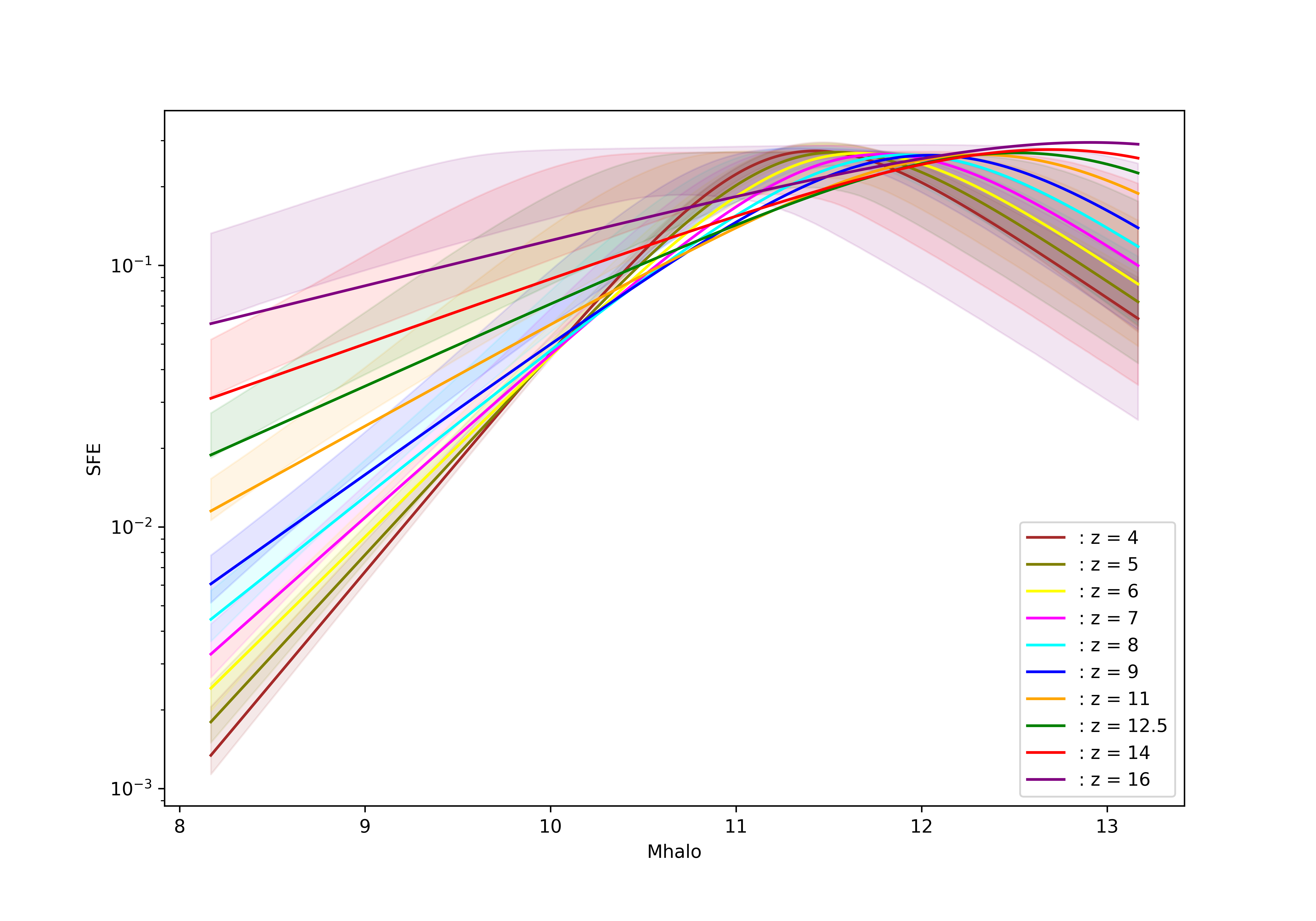}
    \caption{Star formation efficiency versus halo mass evolution for the model with both $\alpha(z)$ and $M_0(z)$ redshift dependence for the another (second) best ranked model, showing evolution in both the low-mass slope and the characteristic mass scale. The x axis halo mass is taken to be in log scale.}
    \label{SFE_m1alpha}
\end{figure}

\begin{figure}
    \centering
    \includegraphics[width=0.7\linewidth]{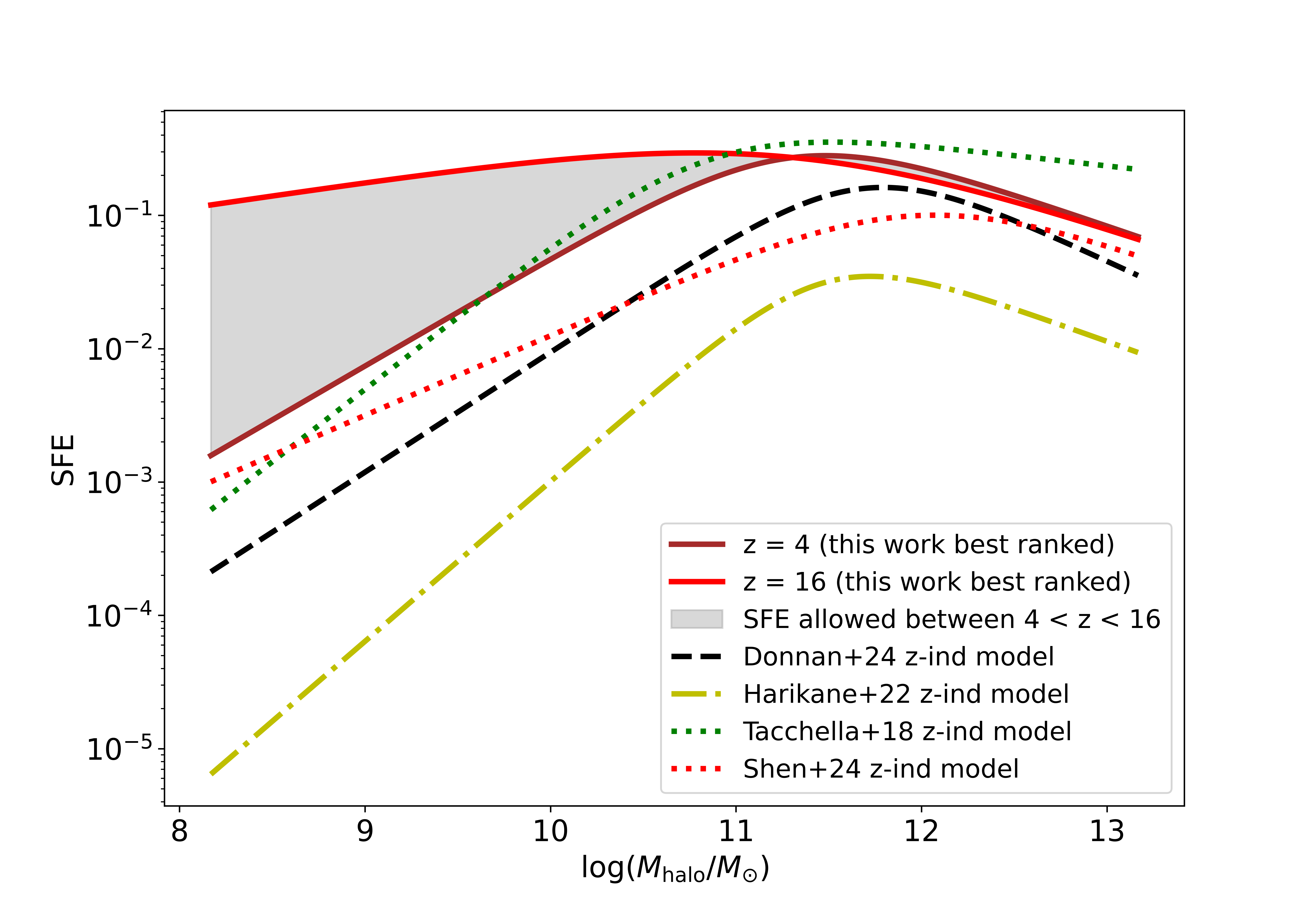}
    \caption{Comparison of star formation efficiency versus halo mass relation between literature and the best ranked model SFE($z$) from this work. SFE - $M_{\rm H}$ relations are taken from \citet{Donnan2024, Tacchella2018, ShenEDE, Harikane2022} and are redshift-independent. The x axis halo mass is taken to be in log scale.}
    \label{SFE_compare}
\end{figure}

\bibliography{main,zot}{}
\bibliographystyle{aasjournalv7}

\end{document}